\documentclass[fleqn,usenatbib]{mnras}

\usepackage{newtxtext,newtxmath}

\usepackage[T1]{fontenc}
\usepackage{ae,aecompl}
\usepackage{multirow}
\usepackage{tabularx}
\usepackage{graphicx}
\usepackage{array}
\usepackage{supertabular}
\usepackage{longtable}
\usepackage[]{url}
\usepackage{graphicx}
\usepackage{comment}
\usepackage{booktabs}
\usepackage{gensymb}
\usepackage{xcolor,color}
\usepackage[caption = false]{subfig}
\usepackage{ulem}
\usepackage[cm]{aeguill}
\renewcommand{\arraystretch}{1.6}
\usepackage{float}
\usepackage{rotating}
\usepackage{amsmath}

\usepackage{xfp}
\usepackage{calc}
\newlength{\mynumberwidth}
\newlength{\myminuswidth}
\def\myscalefactor{\fpeval{\mynumberwidth/\myminuswidth}}
\newcommand{\myminus}{\scalebox{\myscalefactor}[1.0]{$-$}}
\newcommand{\myplus}{\scalebox{\myscalefactor}[1.0]{$+$}}
\DeclareUnicodeCharacter{0301}{é}

\definecolor{darkgreen}{RGB}{0,100,0}

\DeclareRobustCommand{\VAN}[3]{#2}
\let\VANthebibliography\thebibliography
\def\thebibliography{\DeclareRobustCommand{\VAN}[3]{##3}\VANthebibliography}

\title[GRB~230812B and SN2023pel]{Multi-band analyses of the bright GRB~230812B and the associated SN2023pel}

\author[T.~Hussenot-Desenonges et al.]{
\large T.~Hussenot-Desenonges$^{1}$,  T.~Wouters$^{2,3}$,  N.~Guessoum$^{4}$\thanks{Corresponding author},  I.~Abdi$^{4}$,  A.~Abulwfa$^{5}$,   C.~Adami$^{6}$,  J.~F.~Ag\"u\'i~Fern\'andez$^{7}$,  T.~Ahumada$^{8}$, \newauthor \large  V.~Aivazyan$^{9,10}$,  D.~Akl$^{4}$
% ...
}

\begin{document}
\label{firstpage}
\pagerange{\pageref{firstpage}--\pageref{lastpage}}
\maketitle

\begin{abstract}

GRB~230812B is a bright and relatively nearby ($z =0.36$) long gamma-ray burst (GRB) that has generated significant interest in the community and has thus been observed over the entire electromagnetic spectrum. We report over 80 observations in X-ray, ultraviolet, optical, infrared, and sub-millimeter bands from the GRANDMA (Global Rapid Advanced Network for Multi-messenger Addicts) network of observatories and from observational partners. Adding complementary data from the literature, we then derive essential physical parameters associated with the ejecta and external properties (i.e. the geometry and environment) of the GRB and compare with other analyses of this event. We spectroscopically confirm the presence of an associated supernova, SN2023pel, and we derive a photospheric expansion velocity of v $\sim$ 17$\times10^3$ km s$^{-1}$. We analyze the photometric data first using empirical fits of the flux and then with full Bayesian Inference. We again strongly establish the presence of a supernova in the data, with a maximum (pseudo-)bolometric luminosity of $5.75 \times 10^{42}$ erg/s, at $15.76^{+0.81}_{-1.21}$ days (in the observer frame) after the trigger, with a half-max time width of 22.0 days. We compare these values with those of SN1998bw, SN2006aj, and SN2013dx. Our best-fit model favours a very low density environment ($\log_{10}({n_{\rm ISM}/{\rm cm}^{-3}}) = -2.38^{+1.45}_{-1.60}$) and small values for the jet's core angle $\theta_{\rm core} = 1.54^{+1.02}_{-0.81} \ \rm{deg}$ and viewing angle $\theta_{\rm obs} = 0.76^{+1.29}_{-0.76} \ \rm{deg}$. GRB 230812B is thus one of the best observed afterglows with a distinctive supernova bump.

\end{abstract}


\section{Introduction}

Gamma-ray bursts (GRBs) are energetic explosions that, with their afterglows, emit over the entire range of electromagnetic radiation. Typically, they are classified in two categories: "long" (duration), lasting more than 2 seconds in the gamma/X-ray bands, and "short", lasting less than 2 seconds. Long GRBs are widely believed to result from the collapse and explosion of a very massive star, hence they are often referred to as "collapsar" and hypernova. Short GRBs are thought to result from the merger of a neutron star with another neutron star or with a black hole (compact objects).  In both categories, GRBs produce bipolar jets emerging from the newly formed compact object. The jets interact with the surrounding matter and, through shocks, producing "afterglow" emission, first in the X-ray band, then, as the jet slows and weaker shocks occur, UV, optical, IR, and radio emissions. The luminosity of GRB afterglows is moderately correlated with the isotropic prompt-emission (mostly $\gamma$-ray) energy released, $E_{\rm iso}$ \citep{2008ApJ...689.1161G,2009ApJ...701..824N,2010ApJ...720.1513K,2011ApJ...734...96K}. 

Core-collapse GRBs are also associated with an optical/near-infrared supernova (SN), which represents the more isotropic outburst (in addition to the jet) from the central explosive process (more below). GRB~980425/SN1998bw was the first well-documented example of a GRB associated with a supernova, a core-collapse event strongly associated with a (long-duration) burst \citep{Galama_1998, Patat_2001}. The association between SN1998bw and GRB~980425 was first made from the optical spectrum and location of a transient in a spiral arm of the galaxy ESO184-G82 \citep{Galama_1998}. Observations of such cases strengthen the fact that galaxies with strong star formation have greater potential for the occurrence of long gamma-ray bursts \citep{Bloom_2002}.

In addition to this seminal case, a number of similar associations have been observed since then, such as GRB~030329 with SN2003dh (\citealt{Hjorth2003, Stanek2003}). Spectral features of SN2003dh indicated a very massive star origin \citep{Deng_2005}, reinforcing the notion that the GRB resulted from a core-collapse process. These breakthrough observations opened the door for the collection of a significant number of GRB-SN associations, now a well-identified class of astrophysical events. Other thoroughly studied examples include GRB~031203/SN2003lw (\citealt{Malesani2004}), GRB~060218/SN2006aj (\citealt{ferrero2006, pian2006}), GRB~100316D/SN2010bh (\citealt{cano2011}), GRB~120422A/SN2012bz (\citealt{melandri2012, schulze2014}), GRB~130702A/SN2013dx (\citealt{delia2015, Mazzali2021}), GRB~161219B/SN2016jca (\citealt{Ashall_2019}), and GRB~171010A/SN2017htp (\citealt{2019MNRAS.490.5366M}).

Important elements which strengthen the association of supernovae with gamma-ray bursts (such as the above cases) include the broad lines in the object's emission spectrum, which are strongly typical of Type Ic supernova lines, and the association with star-forming galaxies; such indicators provide a coherent scenario for the GRB-SN association. Still, important issues remain to be resolved, including: in which cases does a process (collapsar, merger) produce a long or short GRB, and what are their counterparts (r-process or not?); what powers the central engine in each case (magnetar, radioactive heating, etc.); what kinds of jets are produced, etc. To address such key questions, we need a large sample of GRB events exhibiting multi-band emission with a (supernova/kilonova) bump in the light curve, characteristic lines in the spectrum, rich enough data to give us constraints on the current models (radioactive heating, millisecond magnetar central engine, etc.).

Important parameters that characterize the SNe associated with GRBs include their maximum luminosity (bolometric and in various bands), the time of their peak emission, and the width at half-max of the light curve. Many authors often use the (less physically meaningful) relative brightness factor $k$ (typically compared to SN1998bw) and the (time) "stretch factor" $s$ (that is, a "time width"), also compared to SN1998bw \citep{Cano_2017}. 
A grading system, introduced by \citet{2012grb..book..169H}, became widely adopted for characterizing the strength of a GRB-SN association. The grading ranges from (A), very strong (conclusive spectroscopic evidence), to (E), the weakest associations. In the last 25 years, there have been a dozen cases rated A or A/B \citep{Cano_2017}. For those, the average peak time (in the rest frame) is $\sim 13.2$ days, with a standard deviation of $\sim 2.6$ days (computed from Table 3 of \citealt{Cano_2017}). 

However, the massive star origin for all long GRBs has recently been challenged by the discovery of a few long GRBs (GRB 211211A, GRB 230307A) associated with a kilonova, normally the signature of a binary compact object merger \citep{Rastinejad_2022, Troja_2022, Yang_2022, 2023GCN.33578....1B, 2023GCN.33569....1L}. In addition to these recent associations with kilonovae, there are also nearby long GRBs without a detected bright SN \citep{fynbo2006, Gal-Yam2006, Della_Valle2006, gehrels2006, jin2015}. This evidence then produces a more nuanced picture: while most long GRBs originate in massive star explosions, a few may have a different origin. It is thus crucial to obtain a revised census of the collapsar/merger origin for long GRBs. Events at low redshift ($z \lesssim 0.5$) offer an excellent opportunity to carry out this measurement, as the associated SNe, if present, can be easily detected in photometry and even confirmed spectroscopically with 10m-class telescopes. GRB 230812B provided us with an opportunity to further explore these GRB-Supernova/Kilonova associations.

GRB-SN associations may also be found serendipitously with optical wide-field survey programs \citep{Soderberg_2007} rather than by following bursts and their afterglows. GRB 230812B was initially detected by the \textit{Fermi} Gamma-ray Burst Monitor (GBM - \citealt{2009ApJ...702..791M}), the Gravitational wave high-energy Electromagnetic Counterpart All sky Monitor (GECAM), the AGILE/MCAL instrument (\citealt{2023GCN.34402....1C}), and the Konus-Wind instrument (\citealt{2023GCN.34403....1F}).  This GRB is the most recent event to exhibit a clear SN feature. 

Triggered at $T_0 =$ 18:58:12 UT on 12 August 2023 (GBM trigger 713559497/230812790 - \citealt{2023GCN.34386....1F}), the GRB's light curve in the [10 - 10 000] keV band showed a very bright short pulse with a $T_{90}$ duration (90\% of its fluence at [50, 300] keV) equal to 3.264 $\pm$ 0.091~s \citep{2023GCN.34391....1R}. The GECAM light curve reported a value of $T_{90}$ = 4~s in the [6-1000] keV range \citep{2023GCN.34401....1X}, and Konus  a total time of $\sim$ 20~s in the [20-1200] keV range \citep{2023GCN.34403....1F}, consistent with GBM's value.The \textit{Fermi} Large Area Telescope (LAT) independently detected high energy photons with a maximum of 72 GeV ($T_0 + 30$ s) \citep{2023GCN.34392....1S}.

With the sky localization probability area provided by GBM or LAT \citep{2023GCN.34387....1L,2023GCN.34392....1S}, a series of tiled observations were obtained by the \textit{Neil Gehrels Swift} observatory X-ray telescope (XRT) \citep{2004ApJ...611.1005G}, the Zwicky Transient Facility \citep{2023GCN.34397....1S}, and the Global MASTER-Net \citep{2023GCN.34389....1L}. The X-ray and UV counterpart of GRB~230812B was discovered 7.1 hours after $T_0$ by \textit{Swift}/XRT \citep{2023GCN.34394....1P} and \textit{Swift}/UVOT \citep{2023GCN.34399....1K}. The optical counterpart of GRB~230812B was found by the Zwicky Transient Facility on 2023-08-13 at 03:34:56, 8.5 hours after the GRB trigger time T$_0$ \citep{2023GCN.34397....1S}, and also by  KAIT (the Katzman Automatic Imaging Telescope - \citealt{KAIT2023GCN}), which provided localization with arcsecond accuracy. Simultaneously, the Global MASTER-Net robotic telescopes network reported the optical counterpart at the same location \citep{2023GCN.34396....1L}.

A series of photometric observations across the full electromagnetic spectrum were conducted in the months following the trigger. Among them, we can cite as an example the Multi-purpose InSTRument for Astronomy at Low-resolution spectra-imager (T193/MISTRAL) in optical \citep{2023GCN.34418....1A, 2023GCN.34743....1A, 2023GCN.34762....1A}, the Italian 3.6m TNG telescope in near-infrared, and the Northern extended millimeter array (NOEMA) in radio \citep{2023GCN.34468....1D}. Spectroscopic observations were also conducted in parallel. It led to the measurements of the transient's redshift: $z = 0.360$ \citep{2023GCN.34409....1D}. Twelve days later, observations using OSIRIS+ mounted on the Gran Telescopio Canarias (GTC) showed features in the spectrum characteristic of a GRB-SN event and matched with the spectrum of SN1998bw, indicating, rather conclusively, the presence of a supernova \citep{2023GCN.34597....1A,2023TNSCR2115....1F}. 

Observations with GRANDMA (Global Rapid Advanced Network for Multi-messenger Addicts:  \citealt{GRANDMAO3A,GRANDMAO3B,2022MNRAS.515.6007A,2023ApJ...948L..12K}) observatories started on 2023-08-13T13:34:22, 0.77 days after $T_0$, and lasted for 38 days \citep{2023GCN.34404....1M,2023GCN.34425....1P}. In total, more than 20 professional telescopes and several amateur telescopes imaged the source. 

GRB 230812B being a high-luminosity and (relatively) close-by burst ($z = 0.36$) makes it a very worthwhile target of investigation of the GRB and its afterglow, the SN features, and the correlations between the two. 
To compute distances, absolute magnitudes from apparent magnitudes, etc., we use the \texttt{Planck18} cosmological model from \texttt{astropy} \citep{Planck_2018}; it adopts a flat cosmology with $H_0 = 67.66~\mathrm{km}\cdot\mathrm{s}^{-1}\cdot\mathrm{Mpc}^{-1}$ and $\Omega_\mathrm{m}=0.310$. The observed redshift $z=0.360$ then corresponds to a luminosity distance of $1981 ~\mathrm{Mpc}$; with the fluence $2.52 \times 10^{-4}$ erg~cm$^{-2}$ given by \textit{Fermi}/GBM \citep{2023GCN.34391....1R}, we obtain the total isotropic gamma energy $E_{\gamma, \rm iso} = 1.2 \times 10^{53}$ erg; and with the duration ($T_{90} = 3.26$ s) we get the mean gamma-ray isotropic luminosity $L_{\gamma, \rm iso} = (1+z) E_{\gamma, \rm iso}/T_{90} = 4.9 \times 10^{52}$ erg~s$^{-1}$. This makes this event one of the most luminous GRB-SN events ever recorded.

In this paper, we report observations by the GRANDMA network and its partners of the bright GRB 230812B and the supernova (named SN2023pel) that emerged in the light curve about five days after the burst onset. In \S\ref{data}, we present the observational data from more than two dozen instruments and the photometric methods we use. We also explore properties from the host galaxy (brightness, line of sight extinction). In \S\ref{spectral}, we analyse our multi-epoch spectra from the GRB afterglow to the confirmation of the presence of SN2023pel.  In \S\ref{dataanalysis}, we present the methods we applied in the analysis of the afterglow light curves, using both empirical fits and Bayesian inference. We then present our results on the astrophysical scenarios and processes using different jet structures that best describe the data, and compare SN properties with other GRB-associated supernovae. In \S\ref{conclusion}, we present some general discussion and conclusions.

\section{Observational data}
\label{data}
\subsection{\textit{Swift} XRT, UVOT}
\label{sec:swift_afterglow}

The X-ray light curve ($0.3-10$ keV) of GRB 230812B was acquired from the UK \textit{Swift} Science Data Centre\footnote{\url{https://www.swift.ac.uk/}} \citep{2007A&A...469..379E, 2009MNRAS.397.1177E}. The data were extracted from the Burst Analyser\footnote{\url{https://www.swift.ac.uk/burst_analyser/00021589/}} \citep{2010A&A...519A.102E}, which provides the light curves and spectra of the [0.3-10] keV apparent flux, as well as the unabsorbed flux density at 10 keV in Jansky units. 
For the spectral energy distribution (SED) fitting to measure the dust from the galaxy (see sections below), the [0.3-10 keV] XRT data were grouped by 10 counts/bin using \texttt{grppha}, a subpackage from HEASoft (version 6.31.1), for statistical purposes.
For the other analyses, we performed a re-binning of the unabsorbed light curve at 10 keV by dividing the observations into eight non-continuous time windows. Among these, four windows contained a cluster of observations occurring within an hour or less, while the remaining four had a single data point each. For each cluster, we computed the mean value and standard deviation to produce data points in the light curve for the analysis. These values are reported in Table~\ref{tab:Xandradio}.



We retrieved images taken by the Ultraviolet/Optical Telescope (UVOT, 
\citealt{2005SSRv..120...95R}) from the \textit{Swift} archive\footnote{\url{https://swift.gsfc.nasa.gov}}. The source was imaged using the broadband \textit{white} filter from 0.3 days to 8.2 days. In all the images, we checked the effectiveness of the aspect correction. To address the excess broadening induced by pointing jitter from the aging attitude control system \citep{2023GCN.34633....1S}, a meticulous assessment of an early image was conducted to determine where the source counts merge into the background. To accommodate this, a slightly larger aperture of 7.5 arcseconds was used for the source. All further images show that the source is contained in this aperture. Background measurements were obtained by analyzing an annular region extending from 10 to 22 arcseconds (after a careful background region positioning). The later images were summed to get a good signal-to-noise ratio in the usual way using the {\tt Ftool uvotmaghist}\footnote{\url{https://heasarc.gsfc.nasa.gov/ftools/}}.
We then transformed the Vega magnitudes to AB magnitudes by adding 0.8 mag as is appropriate in \textit{white} \citep{2011AIPC.1358..373B}.

The late-time magnitude upper limits suggest that the host galaxy magnitude is faint, \textit{white} $> 23.2$. We tried deriving a near-UV magnitude for the host galaxy from earlier observations from the \textit{Galex gPhoton} database\footnote{\url{https://github.com/cmillion/gPhoton/blob/master/docs/UserGuide.md}}, but were unsuccessful in avoiding contamination by nearby stars. We eventually chose the magnitude $23.54 \pm 0.84$ in $u$-band from the Sloan Digital Sky Survey DR7 \citep{SDSS7} as an approximation of the \textit{white} contribution, making sure we propagated properly its very conservative error bars through flux subtraction.
The UVOT values, corrected from this constant galaxy flux contribution and from Milky way extinction (see below), are reported in Table~\ref{tab:all_observations}.

\medbreak

\subsection{Optical data set}
\label{sec:post-grbobs}

We conducted simultaneous observations with GRANDMA \citep{GRANDMAO3A}, thanks to its operational platform \texttt{SkyPortal} \citep{Coughlin_2023}, and with associated partners, from less than a day after the trigger time $T_0$ up to 38 days (see Figure~\ref{figureraw}). Details on the observational campaign in the various networks can be found in the Appendix. From the images taken, we successfully extracted the photometry of the source and corrected it from the constant flux contribution of the host galaxy and from absorption by dust along the line of sight. The data set can be found in Table~\ref{tab:all_observations}. Our preliminary analysis of the GRANDMA observations has been reported publicly in the General Coordinates Network (GCN)\footnote{\url{https://gcn.nasa.gov/}} \citep{2023GCN.34404....1M,2023GCN.34425....1P}. 

\begin{figure}
    \centering
    \includegraphics[width=\columnwidth]{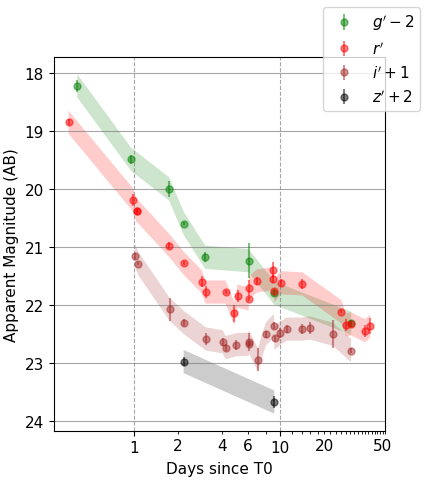}
    \includegraphics[width=\columnwidth]{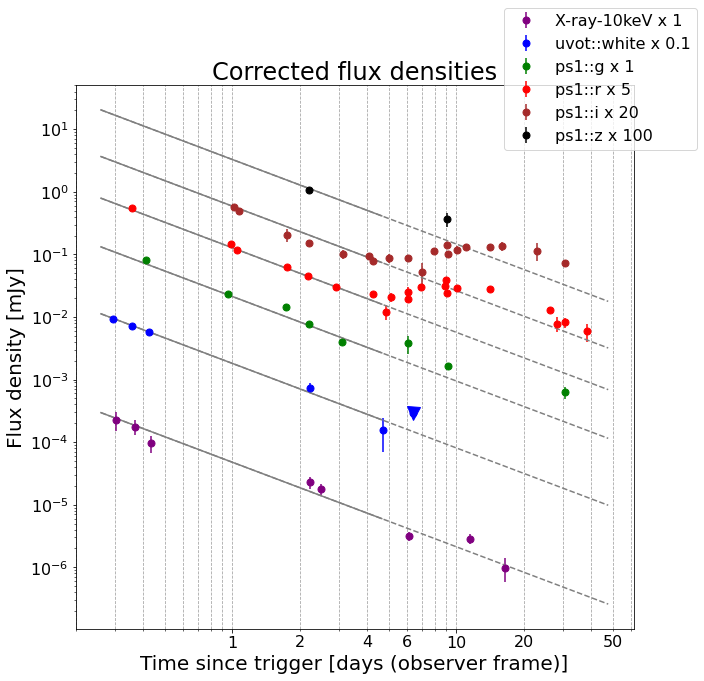}
    \caption{\textbf{Top}: Observations from this work in $g$-, $r$-, $i$-, and $z$-band (observer frame times). Apparent magnitudes before correction by host galaxy flux or for the Milky Way foreground extinction; colored filled regions with arbitrary errorbars 0.2 mag wide have been added to ease the visualization of the light curves. \textbf{Bottom}: Multi-band (X-ray to IR) light curves, corrected from host galaxy flux and the Milky Way foreground extinction. In grey are power-laws fit to the data points up to $T_0 + 5$ days (see section \ref{empiricallc}). }
    \label{figureraw}
\end{figure}

\subsubsection{Photometry}

Prior to photometry, all images were pre-processed in a telescope-specific way with bias and dark subtraction and flat-fielding. We manually masked the regions of the images containing significant imaging artefacts or regions not fully corrected by the pre-processing. Also, we derived astrometric solutions for the images where telescope pipelines did not provide them by using the Astrometry.net service \citep{2010astrometrynet}. 

In order to increase the signal-to-noise ratio of the images, we resampled and coadded individual frames using the {\sc Swarp} software \citep{2010ascl.soft10068B} for sequences of images acquired on the same telescope within a short interval of time.
Then, we performed the forced photometry at the transient position using {\sc STDPipe} \citep{stdpipe}, a set of Python codes for performing astrometry, photometry, and transient detection tasks on optical images, in the same way as \citet{2023ApJ...948L..12K}. 

In order to simplify the analysis and quality checking of the heterogeneous set of images from different telescopes, and to keep track of the results, we created a dedicated web-based application, {\sc STDWeb}\footnote{Accessible at \url{http://stdweb.favor2.info}}, which acts as a web interface to the {\sc STDPipe} library and provides a user-friendly way to perform all steps of its data processing, from masking bad regions to image subtraction, with thorough checking of the intermediate results of every step, and then adjusting the settings in order to acquire optimal photometry results. It also contains some heuristics for the selection of an optimal aperture radius and an optimal selection of reference photometric catalogue, refining the astrometric solution as needed, etc.

Specifically, for the photometry on all images, we used an aperture radius equal to the mean FWHM value estimated over all point-like sources in each image. For photometric calibration, we used the Pan-STARRS DR1 catalogue \citep{Pan-STARRS} for processing the images acquired in filters close to the Sloan system. We used a spatially variable photometric zero-point model represented as a second-order spatial polynomial in order to compensate for the effects of improper flat-fielding, image vignetting, and positionally-dependent aperture correction (e.g. due to PSF shape variations). We first performed the analysis taking into account the linear color term (using $g-r$ for Sloan-like filters) in order to assess how much the individual photometric system of the image deviates from the catalogue one. Then, if the color term is negligible (e.g. smaller than $0.1$), we re-run the analysis of the image without the color term, thus directly deriving the measurement in catalogue photometric system. If the color term is significant, we kept it in the analysis and corrected the measurement using the known color of the transient.

When the signal-to-noise ratio obtained with the forced photometry is below 5, we derive an upper limit for it by multiplying the background noise inside the aperture by 5, and converting this flux value to magnitudes.
For images taken too close to each other (on a logarithmic timescale), we only selected the one with the best signal-to-noise ratio. Images with a sensitivity too low ($>$ 1.5 apparent magnitude brighter than nearby measurements) were excluded from the data analysis.
Images which, after subtraction of the galaxy's constant flux, give a larger error bar than 0.5, were also excluded from our data set for this analysis. 

In parallel, the image reduction for $J$ and $K$ bands was carried out using the {\it jitter} task of the ESO-eclipse package\footnote{\url{https://www.eso.org/sci/software/eclipse/}}.  Astrometry was performed using the 2MASS\footnote{\url{https://irsa.ipac.caltech.edu/Missions/2mass.html}} catalogue.
Aperture photometry was performed using the Starlink PHOTOM package\footnote{\url{http://www.starlink.ac.uk/docs/sun45.htx/sun45.html}}. To minimize any systematic effect, we performed differential photometry with respect to a selection of local isolated and non-saturated reference stars from the UKIDSS\footnote{\url{http://www.ukidss.org/}} survey.

\subsubsection{Host galaxy properties}

The host galaxy of GRB 230812B is SDSS J163631.47+475131.8, with measurements available in SDSS~DR16 \citep{sdssdr16}, but its photometry there is marked as unreliable. The host galaxy's redshift $z=0.36$ was determined through GTC spectroscopic observations of emission lines \citep{2023GCN.34409....1D}. We studied its brightness, both for host flux subtraction and spectral analysis.

\textit{Constant flux from the host at the location of GRB~230812B} -- To better characterize the host galaxy flux, we acquired the data for the GRB position from archival Pan-STARRS DR1 \citep{ps1images} images in $i'$ filters, and from the DESI Legacy Surveys DR10 \citep{legacysurveys} stacked image in $g'$, $r'$ and $z'$ filters. We then performed forced photometry on these images, on the same apertures and with the same parameters as used above for the reduction of the dataset. 
To convert Legacy Survey measurements to the Pan-STARRS photometric system, we estimated the color term\footnote{The color term $C$ here defines the instrumental photometric system through catalogue magnitude and color as $m_{\rm instr} = m_{\rm cat} + C\cdot{\rm color}_{\rm cat}$ and may be fitted during the photometric calibration of the image.} while calibrating these images. For the $g'$ filter, this happened to be negligible, but for $r'$ and $z'$, we used the following equations:
$$r' - 0.11*(g' - r') = 22.73 \pm 0.07$$
$$z' - 0.11*(r' - i') = 22.31 \pm 0.11$$
where the magnitudes $g', r', i', z'$ correspond to the Pan-STARRS system. To extract $r'$ and $z'$, we used the $g'$ values estimated from the Legacy Survey image, and $i'$ values from Pan-STARRS image. 
The results are summarized in Table~\ref{tab:Photometric corrections}.
These host flux contributions were then subtracted from the apparent flux to obtain the transient flux, combining the flux errors from the apparent magnitude and the host contribution to obtain the errors on the host-subtracted flux.

In $J$ and $K$ filters, there are to our knowledge no NIR detections of the host in available survey catalogues. We obtained a deep late-time $J$-band observation at $T_0 + 60$ days using the TNG telescope, finding a magnitude of $20.91 \pm 0.32$ (Vega), \textit{i.e.} $21.82$ (AB). This approximate host galaxy contribution could then be subtracted from the other TNG $J$ images. Unfortunately, no late-time imaging in $K$-band could be performed, so no host contribution could be estimated in this filter.


\begin{table}
\caption{Apparent magnitudes of the host galaxy used for flux subtraction and the Milky Way (MW) extinction in the line of sight in different filters.}
\label{tab:Photometric corrections}
\centering10 - 
\scalebox{1.0}{%
\begin{tabular}{|c|cc|c|}
\hline
Filter &  \multicolumn{2}{c|}{Host galaxy contribution}  & MW Extinction \\
 & Magnitude  & error &  \\
\hline

$u:white$ & 23.54 (AB) & 0.84 & 0.099 \\
$g'$ & 23.78 (AB) & 0.12 & 0.077 \\
$r'$ & 22.83 (AB) & 0.10 & 0.053\\ 
$i'$ & 22.54 (AB) & 0.12 & 0.040\\ 
$z'$ & 22.34 (AB) & 0.12 & 0.029 \\
$J$ & 21.82 (AB) & 0.32 & 0.017 \\
$K$ & - & & 0.007 \\

\hline
\end{tabular}}
\end{table}

\begin{figure}
    \centering
    \includegraphics[width=\linewidth]{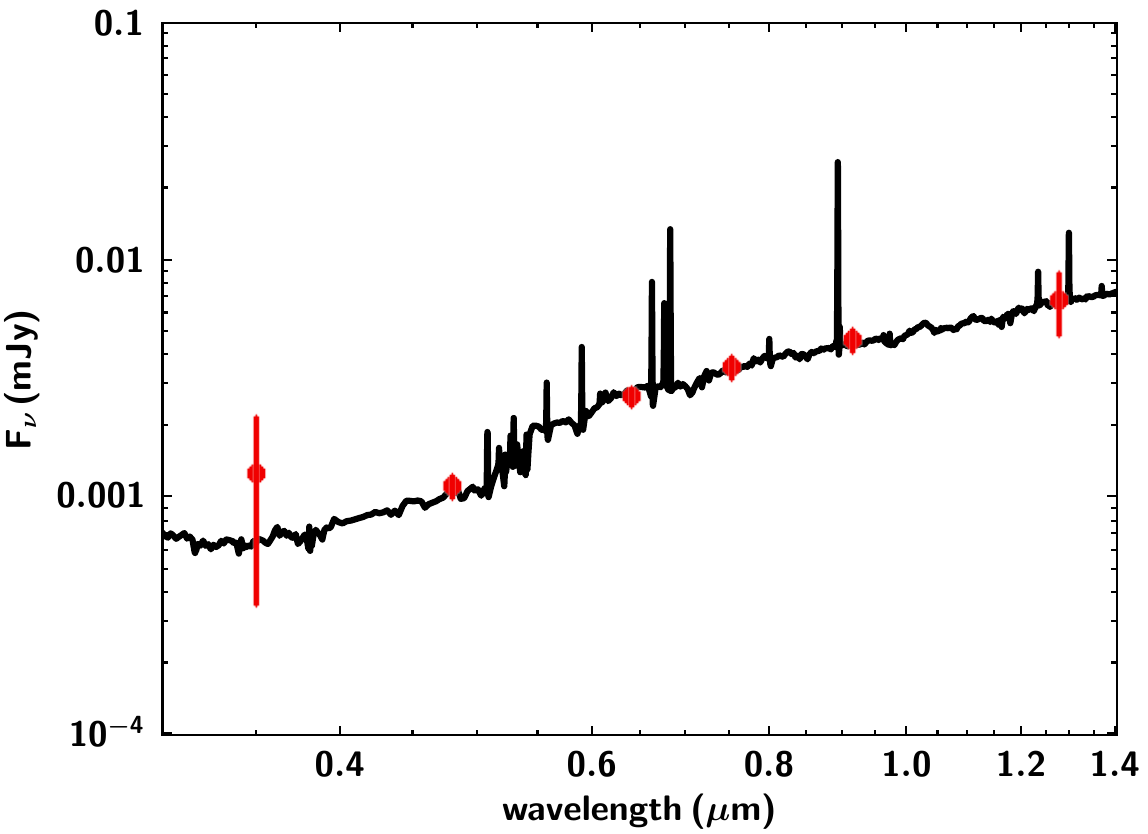}
    \caption{Spectrum of the best-fit host galaxy model in CIGALE, constrained by our estimations in $ugriz$ and $J$ bands.}
    \label{fig:CIGALE}
\end{figure}

\textit{Star formation rate from the host galaxy} -- Using these host flux contributions as approximations for the observed magnitude of the galaxy as a whole, we apply the {\sc CIGALE}\footnote{\url{https://cigale.lam.fr/}} code \citep{cigale} to study the spectral energy distribution (SED) of the galaxy. This analysis constrains the model parameter space to a mass $M = (1.99 \pm 0.54) \times 10^9 M_{\odot}$, a star formation rate (on the last 10 Myr) $SFR = 0.17 \pm 0.07 ~M_{\odot} \text{yr}^{-1}$, and an attenuation $A_V = 0.09 \pm 0.06$ mag. We show the best-fit spectrum in figure \ref{fig:CIGALE}. However, one should keep in mind that we are effectively considering the flux of the host galaxy within the aperture size of the transient (a few arcseconds because of point spread of the instruments; to be compared with the $~5$ kpc/arcsec scale at $z \simeq 0.36$), and are thus missing a fraction of the galaxy, underestimating the flux by an unknown amount that may bias these galaxy parameters. The SFR is especially hard to constrain without more UV data, so its uncertainty provided here is likely underestimated.

\subsubsection{Line of sight extinction}

\textit{Milky Way (MW) extinction} -- We corrected the UV, $griz$, $J$ and $K$ bands from the MW extinction values from \citet{Schlafly2011}, computed along the line of sight by the NED calculator\footnote{\url{http://ned.ipac.caltech.edu/forms/calculator.html}}. These corrections are reported in Table \ref{tab:Photometric corrections}.

\textit{Host galaxy dust extinction} – To estimate the extinction suffered by the afterglow due to the host galaxy dust, we created a spectral energy distribution (SED) from X-ray to optical at two epochs: $T_0$ + 2.2 days, corresponding to the quasi-simultaneity of the \textit{whitegriz} bands, and at $T_0$ + 4 days, to include observations from the \textit{J, K} bands;  as no quasi-simultaneous observation was available at this epoch for \textit{griJ}, the photometric points were estimated through interpolations. We considered the typical extinction curves of MW, Large Magellanic Cloud and Small Magellanic Cloud of \cite{pei1992}, which gave similar results. 

\begin{figure}
    \centering
    \includegraphics[width=\linewidth]{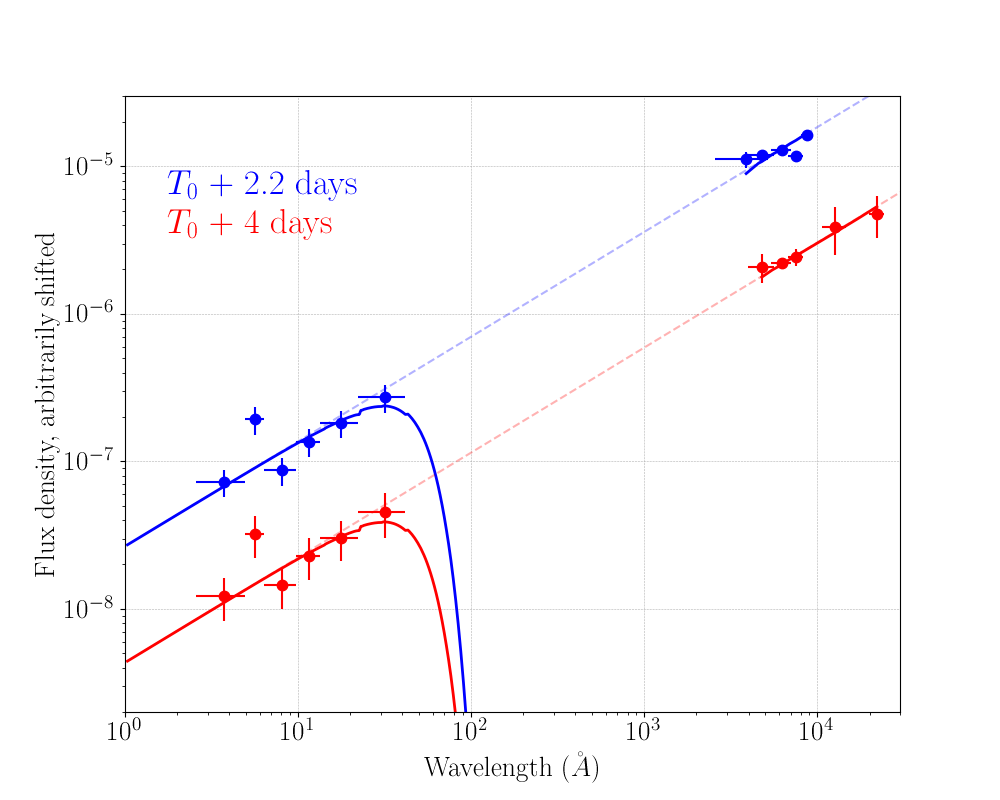}
    \caption{X-ray to NIR SED of the afterglow of GRB~230812B at $T{_0}$ + 2.2 days and $T{_0}$ + 4 days (observer-frame times). The dashed lines correspond to the best fit intrinsic model (single power law). The solid lines illustrate the best fit to the data, including the absorption in the X-ray. The 0.3-10 keV XRT spectrum extracted around $T{_0}$ + 2.2 days has been rescaled and used for the SED at $T{_0}$ + 4 days}
    \label{fig:SEDanalysis}
\end{figure}

We report the results obtained with the average SMC dust extinction law. For each epoch, the intrinsic spectrum was modeled with a single or broken power law using the afterglow theory outlined in \cite{Sari:1997qe}. For the broken power law, the difference in slope between X-ray and NIR wavelengths was set to $\Delta\beta$ = $\beta_X$ - $\beta_o$ = 0.5, which corresponds to the change in slope due to the cooling break. For both epochs, the best fit of the X-ray/NIR SED is obtained with a single power law, and the measured dust extinction $A_V$ is compatible with zero (See Table \ref{tab:SEDepochs1}). The higher uncertainty in $A_V$ for $T{_0}$ + 4 days is due to higher uncertainties in the $J$- and $K$-band observed fluxes. The best fits of the SED at both epochs are shown in Figure \ref{fig:SEDanalysis}.

The $T_0$ + 2.2 days SED constrains best the host galaxy dust extinction as $A_V = 0.0 \pm 0.075$ mag, corresponding to a reddening of $E(B-V) = 0.0 \pm 0.026$ mag for the average SMC model with $R_V = 2.93$ (this constraint is tighter than but compatible with the upper limit $E(B-V) < 0.07$~mag $(3\sigma)$ in \citealt{Gokulpaper}). This is consistent with the {\sc CIGALE} analysis finding a very low global attenuation. We thus chose not to apply any additional extinction correction to the photometric points in Table \ref{tab:all_observations}.

\begin{table}
\centering
\scalebox{1.0}{
\begin{tabular}{|l|l|l|l|}
\hline
 Epoch  & $A_V$ (mag)         & $\beta$        & $\chi^2$ (dof)  \\ \hline
$T_0$ + 2.2 days    & 0.0 $\pm$ 0.075 & 0.710 $\pm$ 0.027 & 19.269 (7)     \\ \hline
$T_0$ + 4 days  & 0.0 $\pm$ 0.185  & 0.712 $\pm$ 0.036 &  4.773 (7)    \\ \hline
\end{tabular}}
\caption{Results of the spectral analysis for both epochs.}
\label{tab:SEDepochs1}
\end{table}

\subsection{Radio}

We also added to our data set two unique submillimeter measurements from NOEMA, takend 3.8 days post $T_0$ : see a brief description of the analysis in \citet{2023GCN.34468....1D}. To complete our multi-wavelength dataset at lower energies, we gathered the published results of radio observations of GRB230812B starting two days after $T_0$ and covering different radio bands from 1 to 15.5 GHz. We use the data from the Arcminute Microkelvin Imager Large-Array \citep{2023GCN.34433....1R}, the Karl G. Jansky Very Large Array \citep{2023GCN.34552....1G, 2023GCN.34735....1C}, and the upgraded Giant Meterwave Radio Telescope \citep{2023GCN.34727....1M}. These data are summarized in Table \ref{tab:Xandradio}. No correction from the host constant flux and extinction were applied to these measurements.

\section{Spectral analysis}
\label{spectral}

\begin{figure}
    \centering
\includegraphics[width=\columnwidth]{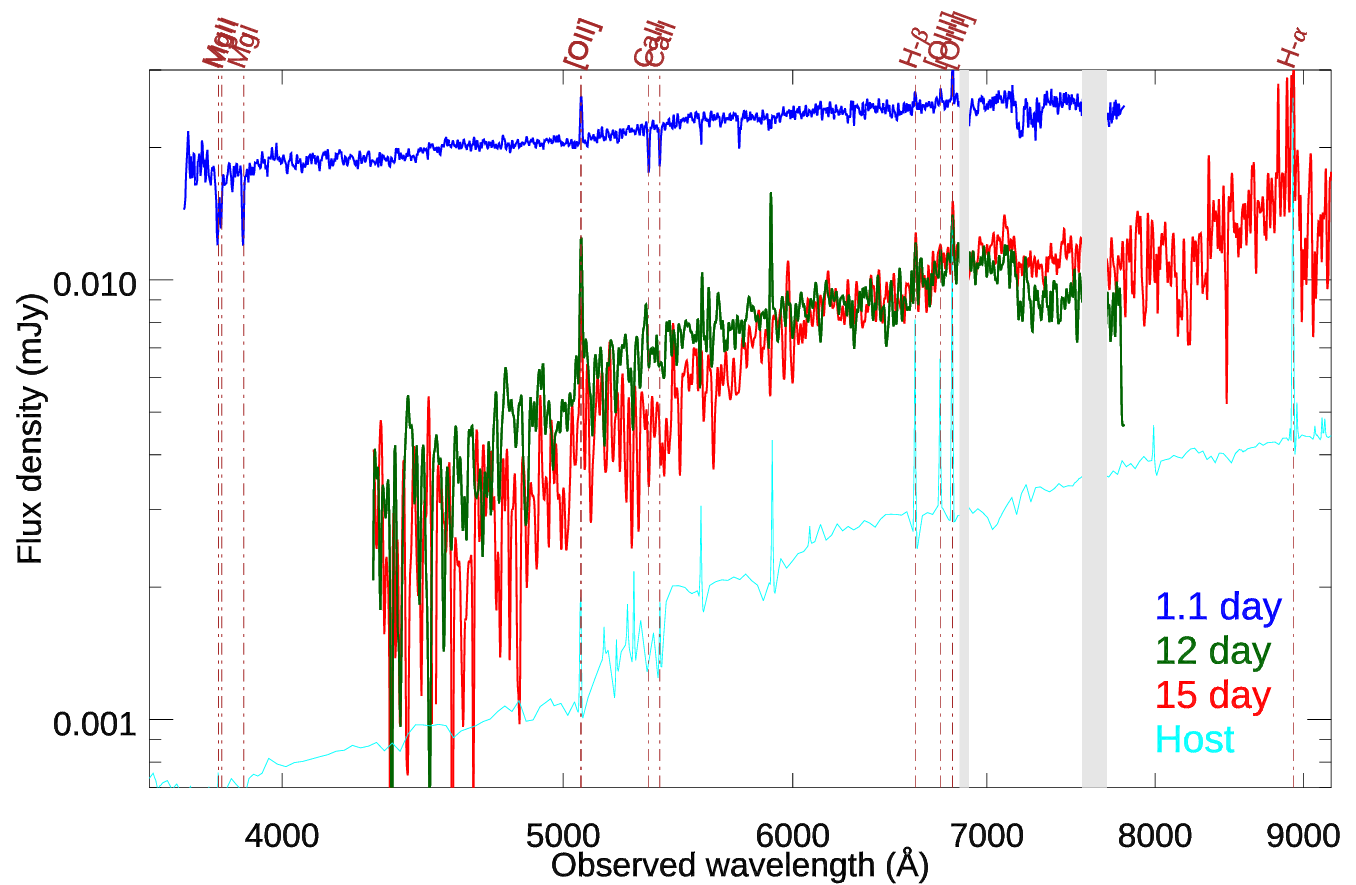}
    \caption{Spectra of GRB\,230812B obtained with OSIRIS+. At 1.1 day the emission was dominated by the afterglow, with a simple powerlaw continuum with absorption lines from the line of sight and emission lines from the host. At 12 and 15 days the supernova component is responsible for most of the emission, with little evolution between the two epochs. We have plotted the host galaxy spectrum derived from the SED fit to understand its contribution to the observations.}
    \label{fig:spectra}
\end{figure}

We performed spectroscopy of the optical counterpart of GRB\,230812B on 3 epochs using OSIRIS+ \citep{2000SPIE.4008..623C} on the 10.4~m Gran Telescopio Canarias (GTC) (see details in the appendix). These spectra, together with the host galaxy model derived from the SED fit are shown in Fig~\ref{fig:spectra}. 

The first epoch was obtained 1.1 day after the GRB, when the strong continuum was dominated by the powerlaw, synchrotron emission of the afterglow. As already mentioned by \citep{2023GCN.34409....1D}, the spectrum shows a strong trace with both emission and absorption lines which we identify as MgII, MgI, CaII, CaI in absorption, and [OII] and [OIII] in emission, at an average redshift of 0.3602$\pm$0.0006, which we identified as the refined redshift of the GRB. The spectral features and their equivalent widths (EW) are displayed in Table~\ref{tab:EWs}. The emission line EWs do not carry much information due to the varying continuum, but the absorption features tell us about the line of sight to the GRB within its own host galaxy. We can calculate the line strength parameter as proposed by \citet{deugarte2012}, which determines the strength of the features as compared to the a large sample of afterglows. The line of sight towards GRB\,230812B displays a line strength parameter of LSP=0.15$\pm$0.16, indicating that the features are just slightly stronger than the average of the sample (percentile 60 of the sample). The only significant difference with respect to the typical GRB spectrum is the relative strength of MgI with respect to MgII. In our case MgI, is relatively strong, implying that the host galaxy of GRB\,230812B is likely to have a low-ionized interstellar environment.

\begin{table}
\centering
\begin{tabular}{|l|l|l|}
\hline
Feature     & Obs. wavelength   &  EW       \\
            & Å                 &   Å       \\
\hline
MgII        & 3801.01         & 2.55$\pm$0.34          \\
MgII        & 3811.40        & 1.76$\pm$0.27          \\
MgI         & 3878.76        & 2.40$\pm$0.29  \\
$[$OII$]$/$[$OII$]$& 5073.53        & -2.26$\pm$0.15          \\
CaII        & 5353.00         & 1.70$\pm$0.16          \\
CaII        & 5401.00        & 1.52$\pm$0.15          \\
CaI         & 5753.43        & 1.23$\pm$0.15          \\
H-beta      & 6614.73     & -0.93$\pm$0.17          \\
$[$OIII$]$      & 6750.73        & -0.74$\pm$0.16          \\
$[$OIII$]$      & 6813.99      & -1.83$\pm$0.15          \\
\hline
\end{tabular}
\caption{Identification and equivalent width of the spectral features observed in the afterglow spectrum.}
\label{tab:EWs}
\end{table}

The other two epochs (12.12 and 15.12 days post $T_0$) show similar, broad features typical of broad line Ic supernovae. The second epoch has a slightly redder continuum, that could be due to the cooling of the ejecta. 
In our analysis, we consider that the contamination by the afterglow continuum is negligible at these epochs.




To analyze the clean SN spectra, we subtracted the contribution from the host galaxy using the host spectrum template that was fit to the host photometry in section 2.2.2. The host subtracted spectra resemble well the ones obtained for SN1998bw at similar rest-frame observing epochs as was earlier noted by \citet{2023GCN.34597....1A}, who identify SN2023pel as a broad line type Ic supernova. 

Furthermore, we use NGSF \citep{2022TNSAN.191....1G} on the host-subtracted spectra to determine the type of SN associated with the burst. For the spectra taken on Aug. 27, the best matches are indeed those of type Ic, the best one being SN2006aj, at phase 2 days (after its peak), with a reduced $\chi^2/dof = 1.74$, followed by SN2002ap, SN2005ek and SN1998bw with $\chi^2/dof = 1.77$, 1.78  and 1.79 respectively. We note that \cite{Gokulpaper} also find SN2002ap and SN1998bw to be good matches to their spectrum.

Additionally, we measured the photospheric velocity of SN2023pel using host-subtracted spectra from GTC. Narrow emission lines and artifacts were first clipped using the IRAF-based routine WOMBAT, and then smoothed the spectra using the the open-source code SESNspectraPCA\footnote{\url{https://github.com/nyusngroup/SESNspectraPCA}}. We measure the velocity of the Fe II line near the SN peak, a proxy for the photospheric velocity of the SN, using SESNspectraLib\footnote{\url{https://github.com/nyusngroup/SESNspectraLib}} \citep{2016ApJ...827...90L,2016ApJ...832..108M}. SESNspectraLib computes the blue-shift of the Fe II 5169~\AA\,line between a normalized SN Ic template and the pre-processed and pre-smoothed SN Ic-BL spectrum.  Since the Fe II feature in a standard SN Ic spectrum is actually a combination of three lines (at 4924~\AA, 5018~\AA, 5169~\AA), one can measure the relative blue shift of the 5169~\AA\,line in a SN Ic-BL spectrum to a normalized SN Ic template of the same phase. The uncertainty on the velocity measurement is calculated by adding the uncertainty of the mean SN Ic template (at a particular phase) in quadrature with the uncertainty on the relative blue-shifted Fe II absorption velocity.
We measure $v_{ph} = 19000 \pm 4000$ km s$^{-1}$ for the spectrum taken on 2023-08-24 and $v_{ph} = 17000 \pm 3000$ km s$^{-1}$ for the spectrum taken on 2023-08-27. 
The velocity we measure is broadly consistent with \cite{Gokulpaper} and with that of the larger population of GRB-SNe at a similar phase \cite{Cano_2017}, for which the average velocity at peak is $v=20000 \pm 8000$ km/s. 


\begin{figure}
    \centering
\includegraphics[width=\columnwidth]{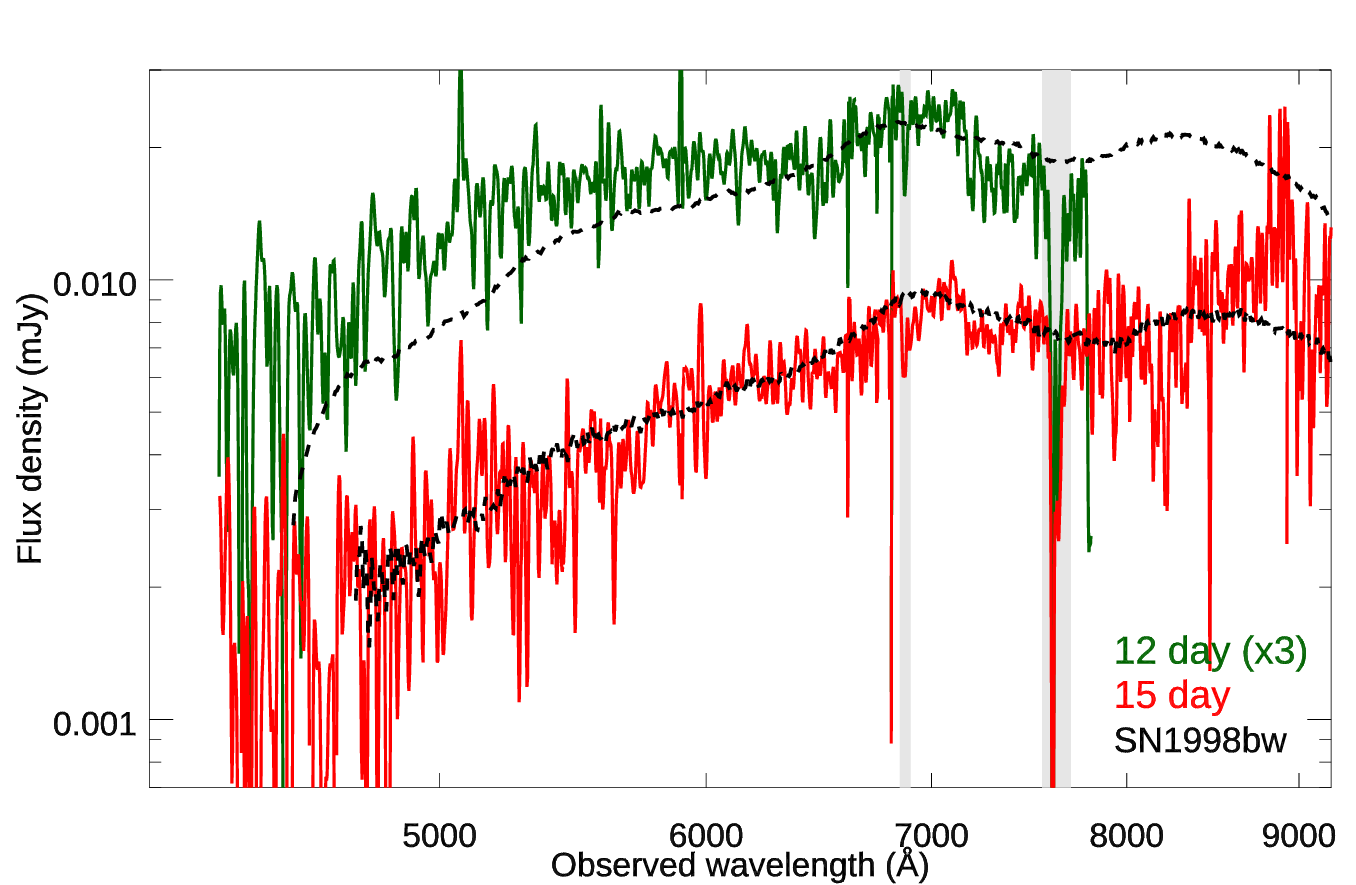}
    \caption{Comparison of the host subtracted spectra of GRB\,230812B at times close to the peak of SN2023pel with spectra of SN1998bw at similar rest frame epochs.}
    \label{fig:spectranohost}
\end{figure}




\section{Multi-wavelength photometric analysis of GRB~230812B and SN2023pel}
\label{dataanalysis}


\subsection{Empirical Light-Curve Analysis}
\label{empiricallc}

\begin{figure}
    \centering
\includegraphics[trim= 100 50 100 0,width=0.9\columnwidth]{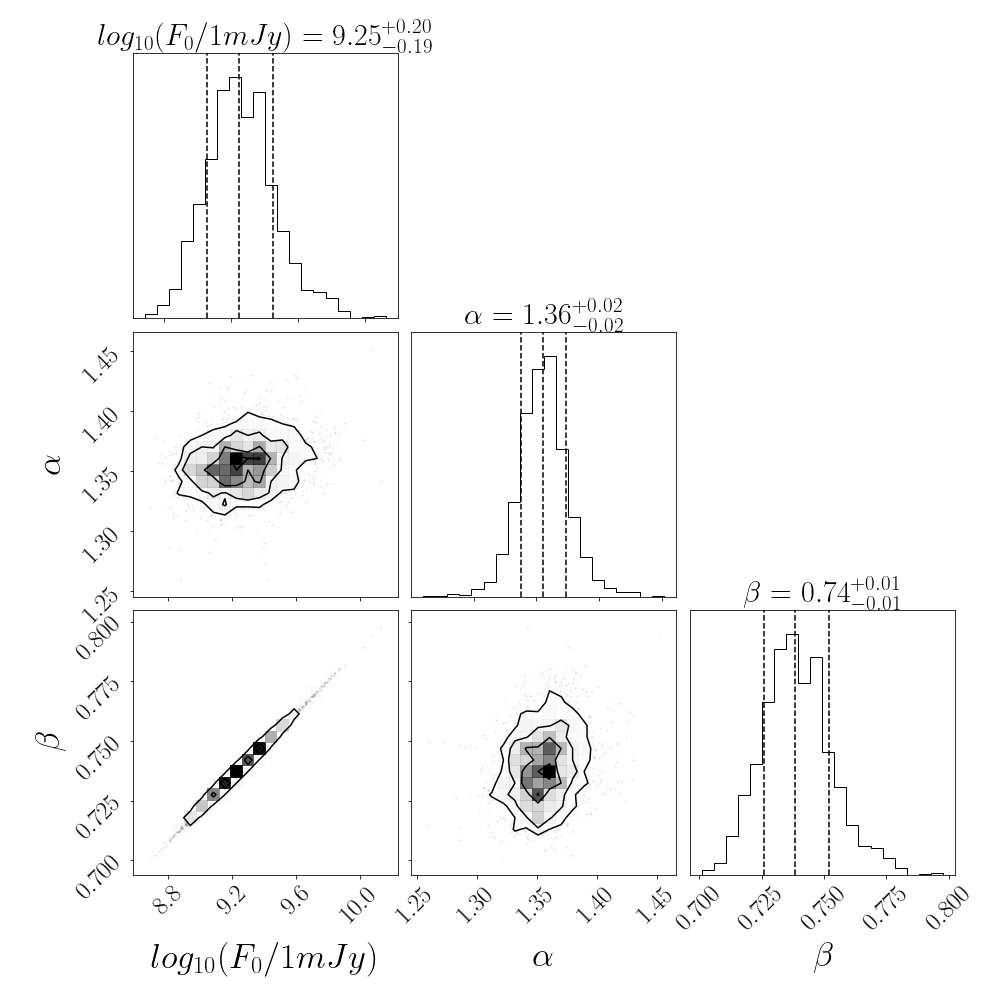}
    \caption{Posteriors of multiband fit of optical afterglow (emission up to 5 days): log of the zero-point flux (at 1~day, 1~Hz, in mJy), $\alpha$ the temporal decay slope and $\beta$ the spectral slope.}
    \label{fig:afterglow_posteriors}
\end{figure}

As a first empirical analysis of the afterglow, we perform a multi-band fit of our data up to 5 days (Figure \ref{figureraw}, bottom), to avoid including the contribution from the emerging supernova. Assuming a power-law function of the form $F_\nu\propto t^{-\alpha}\nu^{-\beta}$, we derive a decay slope of $\alpha = 1.35 \pm 0.02$ and a spectral slope $\beta = 0.74 \pm 0.01$ (Figure \ref{fig:afterglow_posteriors}). We note that these values are almost identical to those obtained by \citet{Gokulpaper} for this GRB: in their work, $\alpha_o = 1.31 \pm 0.02$ and $\beta_o = 0.74 \pm 0.02$. These slopes give an indication of the physical conditions in the GRB's jet (which produces the afterglow through shocks), particularly the electron distribution's index $p$ ($N_e (E) \propto E^{-p}$). 

Using the forward shock model, different assumptions about the afterglow environment lead to different analytical equations and relations between $p$ and $\alpha$ and $\beta$ \citep{Sari:1997qe, 2000ApJ...543...66P}. For instance, a fast-cooling scenario describes a spectral index $\beta = p/2$ leading to an unusual $p = 1.48 \pm 0.03$, but for a slow-cooling scenario, $\beta = (p-1)/2$, which would give a more reasonable $p = 2.48 \pm 0.03$. For the time-decay slope $\alpha$, a uniform external medium gives $\alpha = (3/4)(p-1)$, which means $p = 2.80 \pm 0.04$, while a wind medium gives $\alpha = (3p-1)/4$, $p = 2.14 \pm 0.04$.
The temporal and spectral indices are not satisfied by the same value of $p$, thus, a more sophisticated model (e.g.~jet with structure) is needed, and that is what the \texttt{NMMA} Bayesian inference analysis will undertake.

\subsection{Bayesian Inference for the investigation of the jet structure, SN properties and comparison with state of the art (using \texttt{NMMA})}
\label{NMMA}


The GRB+SN astrophysical scenario of GRB230812B is modeled with the combination of two independent models: the semi-analytic code \texttt{afterglowpy} \citep{vanEerten:2009pa,Ryan:2019fhz} allowing for different structures of the jet, and the \texttt{nugent-hyper} model from \texttt{sncosmo}~\citep{Levan:2004sn} for the supernova. We have previously use the same models on similar \cite{2023arXiv230102049K} and \cite{2023ApJ...948L..12K}

In \texttt{afterglowpy} \citep{vanEerten:2009pa,Ryan:2019fhz}, the thin-shell approximation is used for handling the dynamics of the relativistic ejecta propagating through the interstellar medium, and the angular structure is introduced by dissecting the blast wave into angular elements, each of which evolves independently, including lateral expansion. The analytical descriptions in \cite{Sari:1997qe} are used for the magnetic-field amplification, electron acceleration, and synchrotron emission from the forward shock. The observed radiation is then computed by performing equal-time arrival surface integration.
It should be noted that the model does not account for the presence of a reverse shock or an early coasting phase and does not include inverse Compton radiation. This limits its applicability to the early afterglow of very bright GRBs. In addition, it does not allow to explore a wind-like medium, which may be relevant in a case like GRB230812B. In \texttt{sncosmo}, the supernova modelisation is constructed from observations of the supernova SN1998bw associated with the long GRB~980425. 

We use the Nuclear physics and Multi-Messenger Astronomy framework \texttt{NMMA}~\citep{Dietrich:2020efo,Pang:2022rzc,pang2023updated}\footnote{\url{https://github.com/nuclear-multimessenger-astronomy/nmma}} to evaluate the statistical significance of the different jet structures and provide physical properties of the GRB afterglow and the supernova component\footnote{In order to study and compare the statistical significance between different GRB+SN scenarios in \texttt{NMMA}, we assessed as a first step how statistically favored the GRB+SN scenario (regardless of the jet structure) is compared to a compact object merger scenario. This study is presented in the Appendix~\ref{comparedifferentscenarios}.}.
\texttt{NMMA} uses Bayesian inference that allows us to quantify which theoretical model $\mathbf{M}$ fits the observational dataset $d$ best by computing posterior probability distributions $\mathcal{P}(\vec{\theta}) =  p(\Vec{\theta}| d, \mathbf{M})$. Here $\Vec{\theta}$ denotes the model's parameters. These posteriors are computed via Bayes' theorem:
\begin{equation}
    \mathcal{P}(\vec{\theta}) = \frac{p(d|\Vec{\theta}, \mathbf{M}) p(\Vec{\theta} | \mathbf{M})}{p(d | \mathbf{M})} = \frac{\mathcal{L}(\Vec{\theta}) \pi(\Vec{\theta})}{\mathcal{Z}(d)} \, ,
\end{equation}
where $\mathcal{L}(\Vec{\theta})$, $\pi(\Vec{\theta})$ and $\mathcal{Z}(d)$ are called the likelihood, the prior, and the evidence, respectively. The nested sampling algorithm implemented in \textsc{pymultinest} \citep{2016ascl.soft06005B} is used for obtaining the posterior samples and the evidence.

Assuming \textit{a priori} that the different scenarios considered are equally likely to explain the data, the plausibility of $\mathbf{M}_1$ over $\mathbf{M}_2$ is quantified by the Bayes factor
\begin{equation}
    \mathcal{B} = \frac{p(d | \mathbf{M}_1)}{p(d | \mathbf{M}_2)} \, ,
\end{equation}
with $\mathcal{B} > 1 (\ln \mathcal{B} > 0)$ indicating a preference for $\mathbf{M}_1$, and vice versa.
Given a set of AB magnitude measurements $\{m^{j}_{i}(t_i)\}$ (and the associated statistical uncertainties $\sigma^{j}_{i}$) across different times $\{t_i\}$ and filters $\{j\}$, the likelihood is given by
\begin{equation}
\begin{aligned}
    &\mathcal{L}(\vec{\theta})\\
    &= \prod_{ij}\frac{1}{\sqrt{2\pi ((\sigma^j_i)^2 + (\sigma_{\rm sys})^2)}} \exp\left(- \frac{1}{2}\frac{\left(m^{j}_{i} - m^{j, {\rm est}}_{i}(\vec{\theta})\right)^2}{(\sigma^j_i)^2 + (\sigma_{\rm sys})^2}\right),
\end{aligned}
\end{equation}
where $m^{j, \rm{est}}_{i}(\vec{\theta})$ is the estimated AB magnitude for the parameters $\vec{\theta}$ given different models.
Moreover, as an improvement over \cite{2023arXiv230102049K} and \cite{2023ApJ...948L..12K}, the systematic uncertainty $\sigma_{\rm sys}$ is treated as a free parameter and sampled over during the nested sampling and not kept fixed at a particular value. 
Therefore, the resulting posterior of $\sigma_{\rm sys}$ can also be interpreted as the goodness of fit. The lower the $\sigma_{\rm sys}$, the better the fit, and vice versa.

We have analyzed our full data set (X-ray, UV, optical, IR, and radio)\footnote{With the exception of the $K$ band, for which host flux contributions cannot be computed due to the absence of late-time observations.} with \texttt{NMMA}. All the values quoted in this section are medians with a $95\%$ credible interval as uncertainty. 


\subsubsection{Jet structure}

We vary the jet structure of the GRB to try to characterize or to constrain the jet. To do this, we considered Gaussian (Gauss) and power-law (Power-law) jet structures. Gaussian jets feature an angular dependence $E(\theta_{\rm obs}) \propto \exp(-\theta_{\rm obs}^2/(2\theta_c^2))$ for $\theta_{\rm obs}\leq \theta_w$, with $\theta_w$ being an additional free parameter. A power-law jet features an angular dependence $E(\theta_{\rm obs}) \propto (1 + (\theta_{\rm obs} / \theta_c)^2 / b)^{-b/2}$ for $\theta_{\rm obs} \leq \theta_w$, with $\theta_w$ and $b$ being additional parameters. 
The resulting log Bayes factor $\ln\mathcal{B}$ of Power-law+SN relative to Tophat+SN and Gauss+SN is found to be  $18.360 \pm 0.020$ and $17.356 \pm 0.020$, respectively, demonstrating a preference for the power-law jet.

\begin{figure}
    \hspace*{-0.2cm}
    \includegraphics[width=\columnwidth]{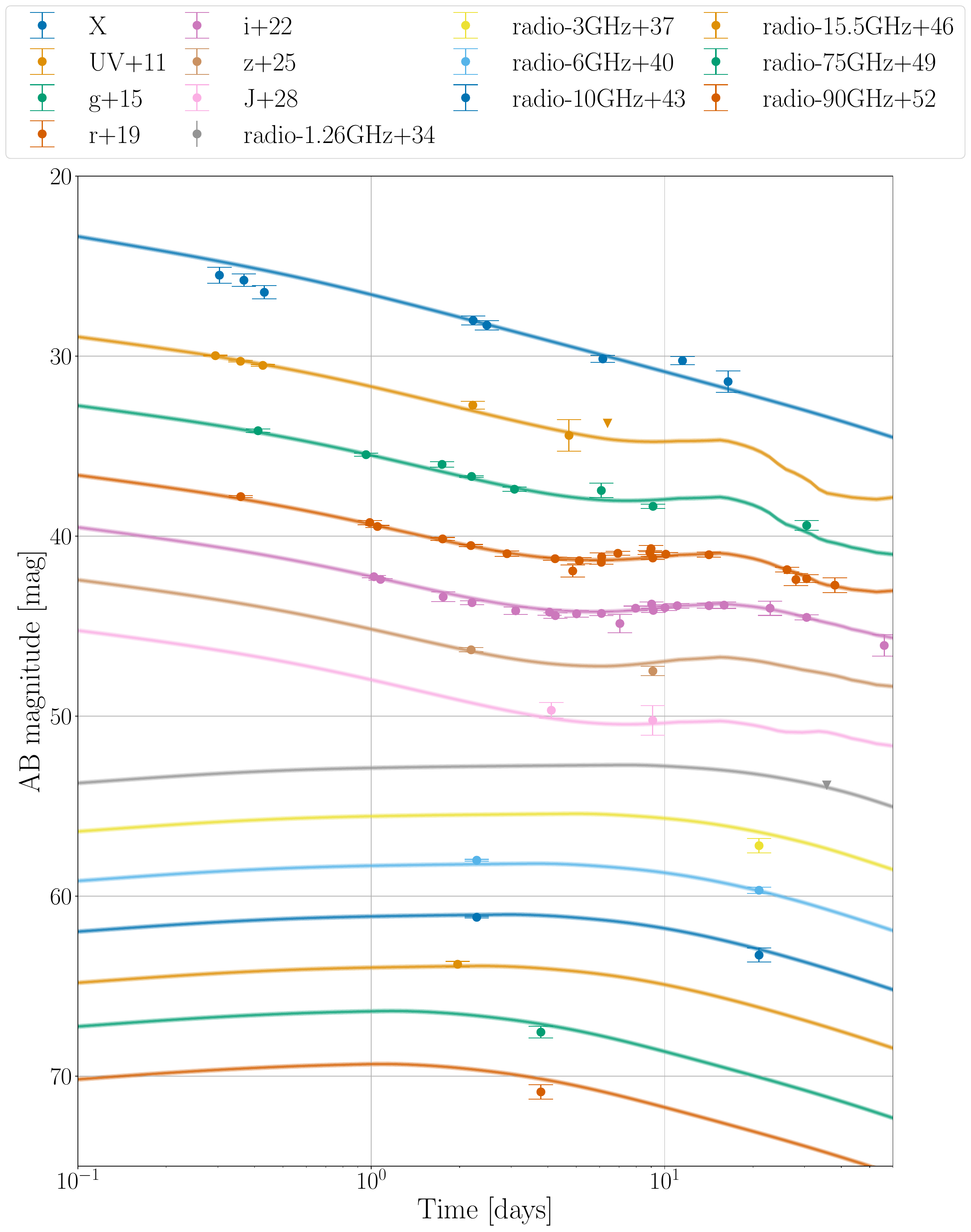}
    \caption{Best-fit light curves of the Power-law+SN model. Datapoints are reported in the observer frame.} 
    \label{fig:bestfit_lcs_nmma_powerlaw}
\end{figure}

\begin{figure*}
    \centering
    \includegraphics[width=2\columnwidth]{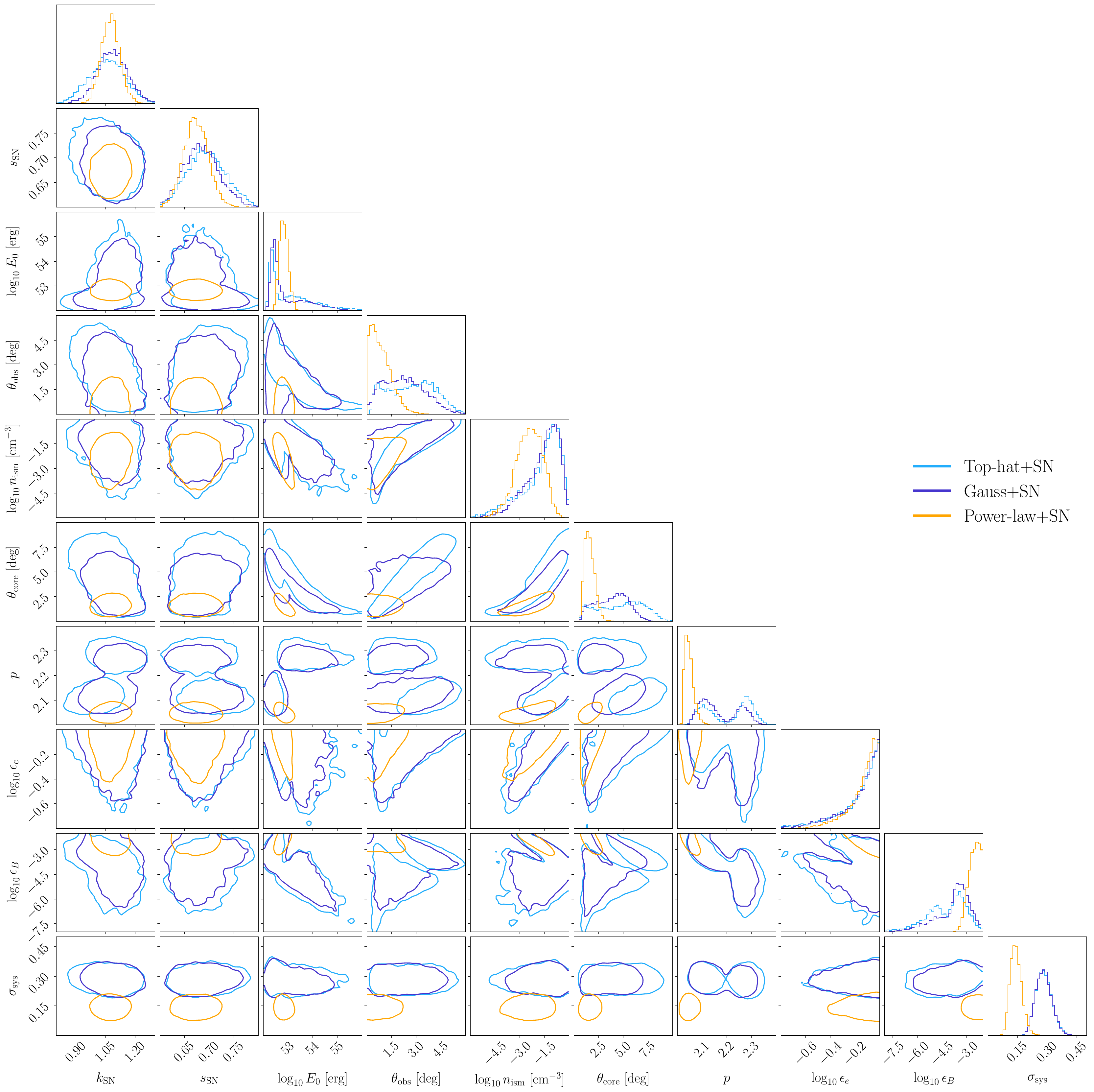}
    \caption{Posterior distribution using different jet models of \texttt{afterglowpy} and \texttt{nugent-hyper} for the supernova component.}
    \label{fig:corner_nmma}
\end{figure*}

The light curve fits of our best-performing model, \textit{i.e.} Power-law+SN, are shown in Figure~\ref{fig:bestfit_lcs_nmma_powerlaw}.
The posterior distributions of the GRB+SN models for all jet structures considered in this work are shown in Figure \ref{fig:corner_nmma} (with corresponding priors displayed in Table \ref{tab:MCMC_paramsNMMA}).
The corresponding best-fit light curves are shown in Figure~\ref{fig:bestfit_lcs_nmma} in the Appendix.


Given the Bayes factors with the interpretation of \citet{Jeffreys1961} and \citet{Kass1995}, one will conclude that the power-law jet is decisively favored against the Gaussian and the top-hat jets. Yet, as previously explained, the models presented in \texttt{afterglowpy} have limitations for early-time GRB afterglow, and the early-time data is also the main source of discriminatory power between different jet structures (as seen in Figure~\ref{fig:bestfit_lcs_nmma}). Thus one can only conclude that there is a preference for power-law jet structure over top-hat and Gaussian jet structures, but it is not a confirmation for detecting such a structure.

In Figure \ref{fig:corner_nmma}, we present the \texttt{NMMA} posteriors for the source parameters, namely the isotropic energy $E_0$, the interstellar medium density $n_{\rm ISM}$, the viewing angle $\theta_{\rm obs}$, the half-opening angle of the jet core $\theta_{\rm core}$, and the microphysical parameters $\{p, \epsilon_e, \epsilon_B\}$ (the power-law index of the electron energy distribution, the fraction of energy in electrons, the fraction of energy in the magnetic field, respectively) using the different jet structure models with SN.

The numerical results for the posteriors and the associated priors can be found in Table \ref{tab:MCMC_paramsNMMA}. For the best-fitting model, namely Power-law+SN, the posterior of $p$ gives $p = 2.04^{+0.04}_{-0.02}$. Moreover, we find $\log_{10}({E_0 / {\rm erg}}) = 52.82^{+0.35}_{-0.31}$ and $\log_{10}({n_{\rm ISM}/{\rm cm}^{-3}}) = -2.38^{+1.45}_{-1.60}$ which is rather low. If such an inferred low-density is not uncommon in long GRBs (GRB 990123 - \citealt{Granot2005}; GRB 090510 - \citealt{corsi2010, joshi2021}; GRB 140515A - \citealt{melandri2015}; GRB 160509A - \citealt{fraija2020}), it remains surprising in this case where the supernova association is strong evidence for a massive progenitor. This may reflect a strong reduction of the progenitor mass loss in the last centuries before the collapse or that the environment had likely been blown away before the jet's interaction with it.   The fractions of energy in the electrons and in the magnetic field are $\epsilon_e = 10^{-0.10^{+0.10}_{-0.29}}$ and  $\epsilon_B = 10^{-2.29^{+1.02}_{-0.94}}$; and the jet's core angle $\theta_{\rm core} = {1.54^{+1.02}_{-0.81}} \ \rm{deg}$ and viewing angle $\theta_{\rm obs} = {0.76^{+1.29}_{-0.76}} \ \rm{deg}$. \\

\subsubsection{Investigation on the X-ray residual}
Figure~\ref{fig:bestfit_lcs_nmma_powerlaw} shows that the best-fitting model has substantial residuals in the X-ray band, especially at earlier times. To further understand this phenomenon, we have performed additional analyses considering data up to 5 days after trigger time, all with the Power-law model, the best-performing GRB model considered. The analyses consider either only the X-ray data or only the UV, optical, and IR (UVOIR) data. The results vary in significance, as demonstrated for instance by the electron energy distribution index $p$. The analysis with the UVOIR dataset gives $p = 2.39^{+0.11}_{-0.14}$, whereas the analysis with only X-ray data gives $p = 2.25^{+0.22}_{-0.27}$. We should, however, note the limitations of this restricted analysis as it results in posterior distributions that are less constrained due to the lower amount of data considered, in the X-ray band, in particular. Moreover, \texttt{afterglowpy} does not include early-time components such as a reverse shock or inverse Compton radiation.
The UVOIR data have a higher weight in the Bayesian analysis due to the higher number of data points in those bands, and since the SN model used here does not support the X-ray band, we can ascribe the high residuals in the X-ray band to a combined effect of the different sizes of the datasets in different filters and a limitation of the models considered in this work.

\subsubsection{Comparison of SN2023pel with other GRB-associated SNe} 

\begin{figure}
    \hspace*{-0.2cm}
    \includegraphics[width=\columnwidth]
    {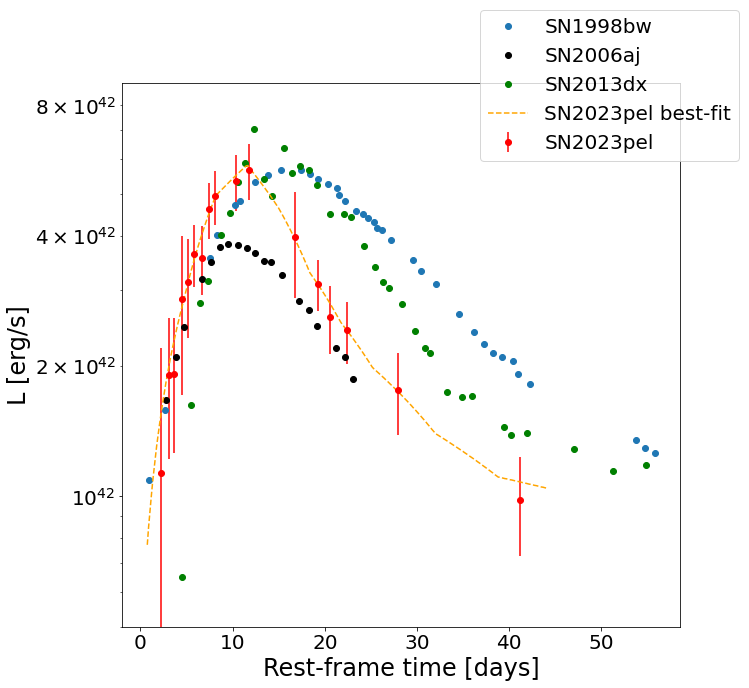}
    \caption{Comparison of pseudo-bolometric ($grizJ$) luminosity (erg/s) of SN2023pel with those of SN1998bw, SN2006aj, and SN2013dx. In orange, the lightcurve computed from the best-fit \texttt{nugent-hyper} model, and in red, the points computed from our observational data after subtraction of the best-fit GRB component.}
    \label{fig:SN-bolo-compare}
\end{figure}

We compare the luminosity of SN2023pel with those of several well-studied GRB-associated supernovae, namely SN1998bw, SN2006aj, and SN2013dx \citep{Mazzali2021}.
To do this, we use the bolometry tool from SNooPy \citep{2011AJ....141...19B}\footnote{\url{https://csp.obs.carnegiescience.edu/data/snpy}} to obtain a bolometric light curve in the $(450 - 1050)$\,nm range.
We grouped our $griz$ and $J$-band data points (see Table~\ref{tab:all_observations}) into 1-day bins. When an optical band has no observed data in a bin, we use the flux from the best-fit model lightcurves as a proxy. We then subtract the bolometric flux from the GRB component of the best-fit model from the total curve in order to get the SN component. Our results are shown in Figure~\ref{fig:SN-bolo-compare}. 
We compare the maximum luminosities, the peak times, and half-max time widths of those four supernovae (SN2023pel, SN1998bw, SN2006aj, and SN2013dx, see Table~\ref{tab:SN-prop-compare}). SN2023pel has a (rest-frame) peak time of 11.6 days, close to that of SN2013dx, but shorter than for SN1998bw and longer than for SN2006aj. It is also consistent with the average peak-time of $13.2 \pm 2.6$ days that \cite{Cano_2017} find for a collection of GRB-SN events. However, SN2023pel notably declines somewhat faster than the other three, especially SN1998bw, with a half-max time width of 16.2 days compared to 36.0 days for the latter.
We also note the differing evolution of the supernova in each band, e.g. the half-max time widths in the $i^\prime$ and $r^\prime$ bands, measured in the source frame, are $17.32^{+1.27}_{-1.23}$ days and $13.30^{+0.97}_{-0.96}$ days, respectively.

\begin{table}
\caption{Comparison of maximum pseudo-bolometric luminosities, (rest-frame) peak times and half-max time widths of SN2023pel, SN1998bw, SN2006aj, and SN2013dx.}
\label{tab:SN-prop-compare}
\centering 
\scalebox{1.0}{%
\begin{tabular}{|c|c|c|c|}
\hline
SN &  L$_{max}$  & peak time & half-max \\
   &  (erg/s)    & (days)    &  time width (days)\\
\hline

SN2023pel & $5.75 \times 10^{42}$ & 11.6 & 16.2 \\
SN1998bw & $5.67 \times 10^{42}$ & 15.3  & 36.0  \\
SN2006aj & $3.80 \times 10^{42}$ & 9.50  & 26.9  \\
SN2013dx & $7.10 \times 10^{42}$ & 12.3  & 25.1  \\

\hline
\end{tabular}}
\end{table}

Additionally, we compute the brightness factor $k_{\rm SN}$ and (time) stretch factor $s_{\rm SN}$ (both relative to SN1998bw) to compare SN2023pel with the results of the recent \citet{Gokulpaper} paper on this supernova. We find $k_{\rm SN} = 1.08^{+0.09}_{-0.10}$ and $s_{\rm SN} = 0.67^{+0.05}_{-0.05}$, respectively. For comparison, \citet{Gokulpaper} find $k_{\rm SN} \approx 0.92$ and $s_{\rm SN} \approx 0.76$, yielding good agreement. Looking at the known correlation between brightness and time-stretch (see Fig 3 in \citealt{Cano_2017}), SN2023pel is the brightest one among events of similar characteristic times, but still compatible with the distribution.

Further investigations have been pursued by \citet{Gokulpaper} to contextualize GRB 230812B/SN 2023pel with
respect to a complete GRB-SN population beyond the three examples mentioned above (see Sec.~4.3 of \citet{Gokulpaper}). Their comparison (using statistical correlations between $L_{\gamma,iso}$ and $k_{SN,1998bw}$, ${L_\gamma,iso}$ and $M_{Ni}$) shows that SN 2023pel is a rather ordinary SN with respect to the overall GRB-SN population, adding more evidence that the central engine and SN powering mechanisms are decoupled in GRB-SN systems.


\begin{table}
    \renewcommand{\arraystretch}{1.1}
    \centering
    \caption{ \texttt{NMMA} - Parameters and prior bounds employed in our Bayesian inferences. We report median posterior values at 95~\% credibility for various physical scenarios and jet structures for the GRB. ``Uniform'' refers to an uniform distribution, and ``LogUniform'' refers to an uniform distribution for the $\log$ of the parameter. $\mathcal{N}(\mu, \sigma^2)$ refers to a Gaussian distribution with mean $\mu$ and variance $\sigma^2$.}
    \label{tab:MCMC_paramsNMMA}
    \begin{turn}{90}
    \resizebox{0.7\paperheight}{!}{
\begin{tabular}{lccccc}
\toprule
Parameter & Prior & Prior range & Tophat+SN & Gauss+SN & Power-law+SN \\
\midrule
(log-) Isotropic afterglow energy $E_0$ [erg] & $\rm Uniform$ & $[47, 57]$ & $52.91^{\myplus2.20}_{\myminus0.78}$ & $52.58^{\myplus2.00\phantom{0}}_{\myminus0.42\phantom{0} }$ & $52.82^{\myplus0.35\phantom{0}}_{\myminus0.31\phantom{0} }$ \\
(log-) Ambient medium's density $n_{\rm ism} $[$\rm{cm}^{-3}$] & $\rm Uniform$ & $[-6, 3]$ & $\myminus1.36^{\myplus1.34}_{\myminus2.94}$ & $\myminus1.38^{\myplus1.30\phantom{0}}_{\myminus2.27\phantom{0} }$ & $\myminus2.38^{\myplus1.45\phantom{0}}_{\myminus1.60\phantom{0} }$ \\
(log-) Energy fraction in electrons $\epsilon_{\mathrm{e}}$ & $\rm Uniform$ & $[-5, 0]$ & $\myminus0.13^{\myplus0.13}_{\myminus0.49}$ & $\myminus0.12^{\myplus0.12\phantom{0}}_{\myminus0.43\phantom{0} }$ & $\myminus0.10^{\myplus0.10\phantom{0}}_{\myminus0.29\phantom{0} }$ \\
(log-) Energy fraction in magnetic field $\epsilon_{\mathrm{B}}$ & $\rm Uniform$ & $[-10, 0]$ & $\myminus4.02^{\myplus1.83}_{\myminus2.65}$ & $\myminus3.56^{\myplus1.59\phantom{0}}_{\myminus2.52\phantom{0} }$ & $\myminus2.29^{\myplus1.02\phantom{0}}_{\myminus0.94\phantom{0} }$ \\
Electron distribution power-law index $p$ & $\rm Uniform$ & $[2.01, 3]$ & $\phantom{0}2.25^{\myplus0.09}_{\myminus0.19}$ & $\phantom{0}2.15^{\myplus0.16\phantom{0}}_{\myminus0.09\phantom{0} }$ & $\phantom{0}2.04^{\myplus0.04\phantom{0}}_{\myminus0.02\phantom{0} }$ \\
Viewing angle $\theta_{\rm obs}$ [degrees] & $\mathcal{N}(0, \theta_{\rm core}^2)$ & -- & $\phantom{0}2.77^{\myplus2.24}_{\myminus2.42}$ & $\phantom{0}2.20^{\myplus2.28\phantom{0}}_{\myminus2.11\phantom{0} }$ & $\phantom{0}0.76^{\myplus1.29\phantom{0}}_{\myminus0.76\phantom{0} }$ \\
Jet core's opening angle $\theta_{\rm core}$ [degrees] & $\rm Uniform$ & $[0.6, 18]$ & $\phantom{0}4.96^{\myplus3.30}_{\myminus4.20}$ & $\phantom{0}3.96^{\myplus2.45\phantom{0}}_{\myminus3.09\phantom{0} }$ & $\phantom{0}1.54^{\myplus1.02\phantom{0}}_{\myminus0.81\phantom{0} }$ \\
``Wing'' truncation angle $\theta_{\rm wing}$ [degrees] & $\rm Uniform$ & $[0.6, 45]$ & -- & $25.18^{\myplus19.77}_{\myminus17.50}$ & $18.25^{\myplus19.07}_{\myminus11.48}$ \\
Power-law structure index $b$ & $\rm Uniform$ & $[0.1, 7]$ & -- & -- & $\phantom{0}1.64^{\myplus0.49\phantom{0}}_{\myminus0.47\phantom{0} }$ \\
Angle ratio $\theta_{\rm obs} / \theta_{\rm core}$ & -- & -- & $\phantom{0}0.56^{\myplus0.23}_{\myminus0.21}$ & $\phantom{0}0.64^{\myplus0.41\phantom{0}}_{\myminus0.57\phantom{0} }$ & $\phantom{0}0.49^{\myplus0.91\phantom{0}}_{\myminus0.49\phantom{0} }$ \\
Supernova boost $k_{\rm SN}$ & $\rm Uniform$ & $[0.01, 100]$ & $\phantom{0}1.06^{\myplus0.18}_{\myminus0.18}$ & $\phantom{0}1.08^{\myplus0.15\phantom{0}}_{\myminus0.15\phantom{0} }$ & $\phantom{0}1.08^{\myplus0.09\phantom{0}}_{\myminus0.10\phantom{0} }$ \\
Supernova stretch $s_{\rm SN}$ & $\rm Uniform$ & $[0.1, 5.0]$ & $\phantom{0}0.69^{\myplus0.08}_{\myminus0.07}$ & $\phantom{0}0.68^{\myplus0.07\phantom{0}}_{\myminus0.07\phantom{0} }$ & $\phantom{0}0.67^{\myplus0.05\phantom{0}}_{\myminus0.05\phantom{0} }$ \\
Systematic error $\sigma_{\rm sys}$ & $\rm LogUniform$ & $[0.01, 2.0]$ & $\phantom{0}0.28^{\myplus0.08}_{\myminus0.08}$ & $\phantom{0}0.28^{\myplus0.08\phantom{0}}_{\myminus0.07\phantom{0} }$ & $\phantom{0}0.14^{\myplus0.06\phantom{0}}_{\myminus0.05\phantom{0} }$ \\
\bottomrule
\end{tabular}
}
\end{turn}
\end{table}

\section{Discussion and Conclusion}
\label{conclusion}


GRB 230812B was a bright and relatively nearby gamma-ray burst that displayed a number of important features: it was accompanied by a luminous supernova, it produced radiation from a high energy of 72 GeV down to radio wavelengths, and was observed for at least a few months since the initial burst, which was detected by several space detectors. Dozens of images and measurements were taken from observatories across the world, including some 80 data points from our GRANDMA network and partner institutions, necessitating not only careful reductions and analyses but also subtractions of backgrounds, host and Milky Way galaxy absorption (dust) and extinction corrections, etc.

With a duration $T_{90}$ of 3.264 $\pm$ 0.091~s (in the [50 - 300] keV band), GRB 230812B falls in the "long" category, thus (in principle) the result of a very massive star's collapse, which produces powerful jets and (oftentimes) a more isotropic supernova, which may be detected several days after the initial burst and afterglow. However, motivated by recent cases indicating that "long" GRBs may sometimes display kilonova characteristics (which are normally associated with "short", merger-type  GRBs) and vice versa ("short" GRBs displaying collapsar-type characteristics), it was worthwhile to analyze this GRB's multi-band emission to see if it is best fit with a supernova or a kilonova, in addition to determining its jet properties, \textit{i.e.} geometry (observed and core angle) and physical parameters (electron and magnetic field energy fractions, etc.).

In a nutshell, our analyses (both photometric and spectral) found a clear confirmation of a supernova, and, using \texttt{NMMA}, a GRB best fit by a high (but not abnormal) total energy $E_0=10^{52.82^{+0.35}_{-0.31}}$ erg. The associated supernova SN2023pel peaked $15.76^{+0.81}_{-1.21}$ days (in the observer frame) after the trigger, consistent with \citet{Gokulpaper} for this supernova and similar to cases of strong GRB-SN associations (\citealt{Cano_2017}). We also plotted pseudo-bolometric light curves for SN2023pel and three other GRB-associated supernovae (SN1998bw, SN2006aj, SN2013dx); we found this new one to have evolved similarly to the others, albeit somewhat faster (especially in decay times).

Our best-fit model also gave a very low ambient density $n_{\rm ISM} = 10^{-2.38^{+1.45}_{-1.60}}$ cm$^{-3}$, similar to a number of previously modeled cases (see the brief discussion and references given above). Further investigations with different models are called for to confirm and understand all these findings. 

Our \texttt{NMMA} framework/simulation also gave best-fit parameter values for the jet's geometry (shape and core and viewing angles) and physical conditions (electron energy distribution index, electron energy fraction, and magnetic field energy fraction). The jet's geometry/shape was best described by the (angular) `power-law' model; the electron energy distribution index $p$ was found to be $\approx 2.1$, which is quite typical; the best-fit magnetic field energy fraction $\epsilon_B$ was $\approx 10^{-2.4}$, also quite typical. However, the electron energy fraction was found to be rather high: $\epsilon_e \approx 0.5 - 1$ ($10^{-0.10^{+0.10}_{-0.29}}$). The jet's core and viewing angles were found to be small: $\theta_{\rm core}={1.54^{+1.02}_{-0.81}} \ \rm{deg}$ and $\theta_{\rm obs}={0.76^{+1.29}_{-0.76}} \ \rm{deg}$, respectively. Despite these atypical values, the parameters still allow an on-axis jet scenario.  

GRB 230812B, bright and relatively close-by, provided us (the GRANDMA network and its partners) the opportunity to perform dozens of observations in UV, optical, near infrared, and sub-millimeter resulting in some 80 high-quality data points. The light curves in optical showed a distinctive supernova bump, SN2023pel, which turned out to be about as bright as the famous SN1998bw. Our spectroscopic analysis determined a photospheric velocity $v_{ph} = 17000 \pm 3000$ km s$^{-1}$ near the peak, and the host-subtracted spectra was best fit by SN2006aj, slightly better than SN2002ap, SN2005ek and SN1998bw. 

The rich data that we have produced, coupled with  data from other groups (\citealt{Gokulpaper}) and facilities, will help explore this event and other GRB-SN associations with additional tools and models. Covering 9 orders of magnitude in frequency, our multi-band analysis presented some information about the jet and the supernova, but further investigations can help confirm or refine our results.

\section*{Acknowledgement}
This work has been coordinated with Mansi Kasliwal and Brad Cenko's group, with whom we shared common developments and visions for time-domain astronomy tools and methods (e.g Skyportal). We thank Gokul Prem Srinivasaragavan in particular for fruitful exchanges on this object. We also thank the anonymous reviewer for constructive comments that have helped improve the paper.

The GRANDMA collaboration thanks its entire network of observatories/observers, all its partners in observations and analyses, and the amateur participants of its Kilonova-Catcher (KNC) program. 

We dedicate this work to D. A. Kann, whose groundbreaking work in the field of GRBs earned him international recognition over the past two decades. Alex, your contributions to the world of GRB science will always be remembered. We deeply miss you and hope you are proud of the way the GRB community carries on your legacy.

GRANDMA has received financial support from the CNRS through the MITI interdisciplinary programs. 
T.W. and P.T.H.P are supported by the research program of the Netherlands Organization for Scientific Research (NWO). 
S.~Antier acknowledges the financial support of the Programme National Hautes Energies (PNHE).
M.~W.~Coughlin acknowledges support from the National Science Foundation with grant numbers PHY-2347628 and OAC-2117997. C.~Andrade and M.~W.~Coughlin were supported by the Preparing for Astrophysics with LSST Program, funded by the Heising Simons Foundation through grant 2021-2975, and administered by Las Cumbres Observatory.
The Egyptian team acknowledges support from the Science, Technology \& Innovation Funding Authority (STDF) under grant number 45779.
S. Karpov is supported by European Structural and Investment Fund and the Czech Ministry of Education, Youth and Sports (Project CoGraDS -- CZ.02.1.01/0.0/0.0/15\_003/0000437).
N.P.M. Kuin is supported by the UKSA \textit{Swift} operations grant.
N.~Guessoum, D.~Akl, and I.~Abdi acknowledge support from the American University of Sharjah (UAE) through the grant FRG22-C-S68.
M. Ma\v{s}ek is supported by the LM2023032 and LM2023047 grants of the Ministry of Education of the Czech Republic.
J.~Mao is supported by the National Key R \& D Program of China (2023YFE0101200), the Yunnan Revitalization Talent Support Program (YunLing Scholar Award), and NSFC grant 12393813. 
D.~B.~Malesani acknowledges support from the European Research Council (ERC) under the European Union's research and innovation program (ERC Grant HEAVYMETAL No.~101071865).
This research has also made use of the MISTRAL database, based on observations made at Observatoire de Haute Provence (CNRS), France, with the MISTRAL spectro-imager, and operated at CeSAM (LAM), Marseille, France.
The GRB OHP observing team is particularly grateful to J\'erome Schmitt for the major role he has played in the development and operations of the MISTRAL instrument at the T193 telescope.
GRANDMA thanks amateur astronomers for their observations: M. Serreau, K. Francois, S. Leonini, J. Nicolas, M. Freeberg, M. Odeh.
J.-G.D. acknowledge financial support from the Centre National d'\'Etudes Spatiales (CNES).
J.-G.D. is supported by a research grant from the Ile-de-France Region within the framework of the Domaine d'Int\'erêt Majeur-Astrophysique et Conditions d'Apparition de la Vie (DIM-ACAV).
The work of F. Navarete is supported by NOIRLab, which is managed by the Association of Universities for Research in Astronomy (AURA) under a cooperative agreement with the National Science Foundation.
The Kilonova-Catcher program is supported by the IdEx Universit\'e de Paris Cit\'e, ANR-18-IDEX-0001 and Paris-Saclay, IJCLAB. 
This work is based on observations carried out under project number S23BG with the IRAM NOEMA interferometer. IRAM is supported by INSU/CNRS (France), MPG (Germany) and IGN (Spain).
Partly based on observations made with the Gran Telescopio Canarias (GTC), installed at the Spanish Observatorio del Roque de los Muchachos of the Instituto de Astrofísica de Canarias, on the island of La Palma.
DBM acknowledges support from the European Research Council (ERC) under the European Union's research and innovation program (ERC Grant HEAVYMETAL No.~101071865). Partly based on observations made with the Nordic Optical Telescope, owned in collaboration by the University of Turku and Aarhus University, and operated jointly by Aarhus University, the University of Turku and the University of Oslo, representing Denmark, Finland and Norway, the University of Iceland and Stockholm University at the Observatorio del Roque de los Muchachos, La Palma, Spain, of the Instituto de Astrofisica de Canarias. The Cosmic Dawn Center is supported by the Danish National Research Foundation. This work is based on observations collected at the Centro Astron\'omico Hispano en Andaluc\'ia (CAHA) at Calar Alto, operated jointly by Junta de Andaluc\'ia and Consejo Superior de Investigaciones Científicas (IAA-CSIC) (Program code : 23B-2.2-24, PI  Ag\"u\'i Fern\'andez, J. F.). 
This work used Expanse at the San Diego Supercomputer Cluster through allocation AST200029 -- "Towards a complete catalog of variable sources to support efficient searches for compact binary mergers and their products" from the Advanced Cyberinfrastructure Coordination Ecosystem: Services \& Support (ACCESS) program, which is supported by National Science Foundation grants \#2138259, \#2138286, \#2138307, \#2137603, and \#2138296. W. Corradi and N. Sasaki whish to thank Laboratório Nacional de Astrofísica - LNA and OPD staff, Universidade do Estado do Amazonas - UEA and the Brazilian Agencies CNPq and Capes. 
\includegraphics[width=\linewidth]{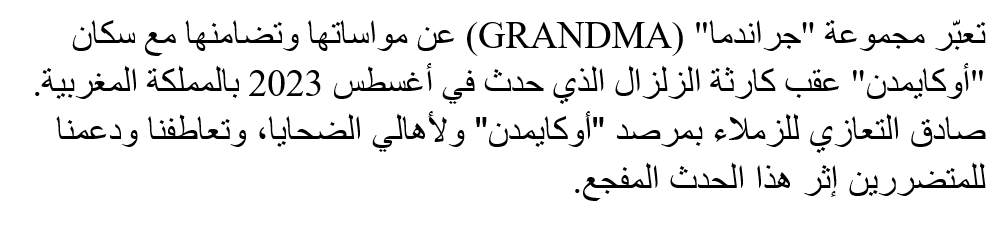}

\section*{Data availability}

Images and raw data are available upon request.

\bibliography{references}{}
\bibliographystyle{mnras}
\appendix

\section*{Affiliations}

$^{1}$ IJCLab, Univ Paris-Saclay, CNRS/IN2P3, Orsay, France\\ 
$^{2}$ Nikhef, Science Park 105, 1098 XG Amsterdam, The Netherlands\\ 
$^{3}$ Institute for Gravitational and Subatomic Physics (GRASP), Utrecht University, Princetonplein 1, 3584 CC Utrecht, The Netherlands\\ 
$^{4}$ Physics Department, American University of Sharjah, Sharjah, UAE\\ 
$^{5}$ National Research Institute of Astronomy and Geophysics (NRIAG), 1 El-marsad St., 11421 Helwan, Cairo, Egypt\\ 
$^{6}$ Aix Marseille Univ, CNRS, CNES, LAM Marseille, France\\ 
$^{7}$ Instituto de Astrof\'isica de Andaluc\'ia, Glorieta de la Astronom\'ia s/n, 18008 Granada, Spain\\ 
$^{8}$ Cahill Center for Astrophysics, California Institute of Technology, Pasadena CA 91125, USA\\ 
$^{9}$ E. Kharadze Georgian National Astrophysical Observatory, Mt. Kanobili, Abastumani, 0301, Adigeni, Georgia\\ 
$^{10}$ Samtskhe-Javakheti State University, Rustaveli Str. 113, Akhaltsikhe, 0080, Georgia\\ 
$^{11}$ School of Physics and Astronomy, University of Minnesota, Minneapolis, Minnesota 55455, USA\\ 
$^{12}$ Observatoire de la C\^ote d'Azur, Universit\'e C\^ote d'Azur, Boulevard de l'Observatoire, 06304 Nice, France\\ 
$^{13}$ INAF–Osservatorio Astronomico di Brera, Via E. Bianchi 46, 23807 Merate (LC), Italy\\ 
$^{14}$ Astronomical Observatory of Taras Shevchenko National University of Kyiv, Observatorna Str. 3, Kyiv, 04053, Ukraine\\ 
$^{15}$ CEICO, Institute of Physics of the Czech Academy of Sciences, Na Slovance 1999/2, CZ-182 21, Praha, Czech Republic\\ 
$^{16}$ Centro Astron\'omico Hispano en Andaluc\'ia, Observatorio de Calar Alto, Sierra de los Filabres, 04550 G\'ergal, Almer\'ia, Spain\\ 
$^{17}$ Laboratoire J.-L. Lagrange, Universit\'e de Nice Sophia-Antipolis, CNRS, Observatoire de la C\^ote d'Azur, 06304 Nice, France\\ 
$^{18}$ National Observatory of Athens, Institute for Astronomy, Astrophysics, Space Applications and Remote Sensing, Greece\\ 
$^{19}$ Institut de Radioastronomie Millim\'etrique\\ 
$^{20}$ Universit\`a degli Studi dell'Insubria, Dipartimento di Scienza e Alta Tecnologia, Via Valleggio 11, 22100 Como, Italy\\ 
$^{21}$ Department of Physics and Earth Science, University of Ferrara, via Saragat 1, I-44122 Ferrara, Italy\\ 
$^{22}$ INFN, Sezione di Ferrara, via Saragat 1, I-44122 Ferrara, Italy\\ 
$^{23}$ INAF, Osservatorio Astronomico d'Abruzzo, via Mentore Maggini snc, I-64100 Teramo, Italy\\ 
$^{24}$ Ulugh Beg Astronomical Institute, Uzbekistan Academy of Sciences, Astronomy Str. 33, Tashkent 100052, Uzbekistan\\ 
$^{25}$ Department of Physics \& Astronomy, Louisiana State University, Baton Rouge, LA 70803, USA\\ 
$^{26}$ Astrophysics Science Division, NASA Goddard Space Flight Center, 8800 Greenbelt Rd, Greenbelt, MD 20771, USA\\ 
$^{27}$ Joint Space-Science Institute, University of Maryland, College Park, MD 20742, USA\\ 
$^{28}$ Laborat\'orio Nacional de Astrof\'isica - LNA, 
Rua Estados Unidos, 154 Itajub\'a - MG CEP 37504-364  Brazil \\ 
$^{29}$ Sorbonne Universit\'e, CNRS, UMR 7095, Institut d'Astrophysique de Paris, 98 bis bd Arago, 75014 Paris, France\\ 
$^{30}$ Institute for Physics and Astronomy, University of Potsdam, Haus 28, Karl-Liebknecht-Str. 24/25, 14476 Potsdam, Germany\\ 
$^{31}$ Max Planck Institute for Gravitational Physics (Albert Einstein Institute), Am M\"uhlenberg 1, 14476 Potsdam, Germany\\ 
$^{32}$ Aix Marseille Univ, CNRS/IN2P3, CPPM, Marseille, France\\ 
$^{33}$ Universit\'e Paris Cit\'e, CNRS, Astroparticule et Cosmologie, F-75013 Paris, France\\ 
$^{34}$ Universit\'e Paris-Saclay, Universit\'e Paris Cit\'e, CEA, CNRS, AIM, 91191, Gif-sur-Yvette, France\\ 
$^{35}$ KNC, AAVSO, Hidden Valley Observatory (HVO), Colfax, WI, USA; iTelescope, New Mexico Skies Observatory, Mayhill, NM, USA\\ 
$^{36}$ Cosmic Dawn Center (DAWN), Denmark\\ 
$^{37}$ Niels Bohr Institute, University of Copenhagen, Jagtvej 128, 2200 Copenhagen N, Denmark\\ 
$^{38}$ N.~Tusi Shamakhy Astrophysical Observatory, Ministry of Science and Education, settl.~Y.~Mammadaliyev, AZ 5626, Shamakhy, Azerbaijan\\ 
$^{39}$ School of Physics and Astronomy, University of Minnesota, Minneapolis, MN 55455, USA\\ 
$^{40}$ Xinjiang Astronomical Observatory (XAO), Chinese Academy of Sciences, Urumqi, 830011, People's Republic of China\\ 
$^{41}$ INAF–Osservatorio Astronomico di Capodimonte, Salita Moiariello 16, I-80131 Napoli, Italy\\ 
$^{42}$ DARK, Niels Bohr Institute, University of Copenhagen, Jagtvej 128, 2200 Copenhagen N, Denmark\\ 
$^{43}$ Division of Physics, Mathematics, and Astronomy, California Institute of Technology, Pasadena, CA 91125, USA\\ 
$^{44}$ IRAP, Universit\'e de Toulouse, CNRS, UPS, 14 Avenue Edouard Belin, F-31400 Toulouse, France\\ 
$^{45}$ Universit\'e Paul Sabatier Toulouse III, Universit\'e de Toulouse, 118 Route de Narbonne, 31400 Toulouse, France\\ 
$^{46}$ Mullard Space Science Laboratory, University College London, Holmbury St. Mary Dorking, RH5 6NT, UK\\ 
$^{47}$ Montarrenti Observatory, S. S. 73 Ponente, I-53018 Sovicille, Siena, Italy\\ 
$^{48}$ Yunnan Observatories, Chinese Academy of Sciences, 650011 Kunming, Yunnan Province, People's Republic of China\\ 
$^{49}$ Department of Astrophysics/IMAPP, Radboud University, 6525 AJ Nijmegen, The Netherlands\\ 
$^{50}$ FZU - Institute of Physics of the Czech Academy of Sciences, Na Slovance 1999/2, CZ-182 21, Praha, Czech Republic\\ 
$^{51}$ Center for Astronomical Mega-Science, Chinese Academy of Sciences, 20A Datun Road, Chaoyang District, 100012 Beijing, People's Republic of China\\ 
$^{52}$ Key Laboratory for the Structure and Evolution of Celestial Objects, Chinese Academy of Sciences, 650011 Kunming, People's Republic of China\\ 
$^{53}$ INAF-Osservatorio Astronomico di Roma, Via di Frascati 33, I-00040, Monte Porzio Catone (RM), Italy\\ 
$^{54}$ Institute of Astronomy and NAO, Bulgarian Academy of Sciences, 72 Tsarigradsko Chaussee Blvd., 1784 Sofia, Bulgaria\\ 
$^{55}$ SOAR Telescope/NSF's NOIRLab Avda Juan Cisternas 1500, 1700000, La Serena, Chile\\ 
$^{56}$ GEPI, Observatoire de Paris, Universit\'e PSL, CNRS, 5 place Jules Janssen, 92190 Meudon, France\\ 
$^{57}$ University of Messina, Mathematics, Informatics, Physics and Earth Science Department, Via F.D. D'Alcontres 31, Polo Papardo, 98166, Messina, Italy\\ 
$^{58}$ Physics Department, Tsinghua University, Beijing, 100084, People's Republic of China\\ 
$^{59}$ Universit\'e de Strasbourg, CNRS, IPHC UMR 7178, 67000 Strasbourg, France\\ 
$^{60}$ Kavli Institute for Astrophysics and Space Research, Massachusetts Institute of Technology, 77 Massachusetts Ave, Cambridge, MA 02139, USA\\ 
$^{61}$ Soci\'et\'e Astronomique de France, Observatoire de Dauban, FR 04150 Banon, France\\ 
$^{62}$ Astronomy and Space Physics Department, Taras Shevchenko National University of Kyiv, Glushkova Ave., 4, Kyiv, 03022, Ukraine\\ 
$^{63}$ National Center Junior Academy of Sciences of Ukraine, Dehtiarivska St. 38-44, Kyiv, 04119, Ukraine\\ 
$^{64}$ Main Astronomical Observatory of National Academy of Sciences of Ukraine, 27 Acad. Zabolotnoho Str., Kyiv, 03143, Ukraine\\ 
$^{65}$ Department of Astronomy, University of Maryland, College Park, MD 20742, USA\\ 
$^{66}$ Department of Physics, Eastern Illinois University, Charleston, IL 61920, USA\\ 
$^{67}$ University of Leicester, Dept. of Physics and Astronomy, University Road, Leicester, LE1 7RH, United Kingdom\\ 
$^{68}$ Astronomical Institute of the Czech Academy of Sciences (ASU-CAS), Fricova 298, Ond\v rejov, 251 65, Czech Republic\\ 
$^{69}$ National University of Uzbekistan, 4 University Str., Tashkent 100174, Uzbekistan\\ 
$^{70}$ University of Catania, Department of Physics and Astronomy. Via Santa Sofia 64, 95123, Catania. Italy\\ 
$^{71}$ Beijing Planetarium, Beijing Academy of Science and Technology, Beijing, 100044, People's Republic of China\\ 
$^{72}$ Department of Astronomy, School of Physics, Huazhong University of Science and Technology, Wuhan, 430074, China

\section{Photometric observations details}

In this section, we detail observations for GRB~230812B by GRANDMA and associated partners. The observations of the optical afterglow of GRB 230812B started on 2023-08-13T13:34:22 UTC, 18.5 hours after the trigger by the \textit{Fermi} Gamma-ray Burst Monitor (GBM), with the GMG 2.4-meter telescope, located at the Lijiang station of Yunnan Observatories. In the context of GRANDMA, this first observation was conducted after the GRANDMA collaboration decided to follow up on this GRB, which goes beyond its standard gravitational-wave follow-up program, 12\,hr after the trigger time. We measured a magnitude of $19.9 \pm 0.1$ in the $R$ band. The TAROT telescopes and other automated systems were inactive during that period.  

The full observational campaign lasted 38 days and ended with observations performed by the 2-meter at Observatoire de Haute Provence. While we took images in $V$, $R$, $I$, $g'$, $r'$, $i'$ and $z'$ bands, we use for this work only data in $g'$, $r'$, $i'$ and $z'$ bands for extracting the physical properties of the event. We however computed the synthetic light curves in $R$, $I$ from the \texttt{NMMA} best-fit parameters constrained from the Xray+UV+$griz$+radio analysis, and confirmed their consistency with our data sets.

Below, in sequence, we here provide the start time (relative to T$_0$) of the first observation for each telescope and the filters/bands used during the entire campaign: GMG (0.78 d in $R$, $g^\prime$, $r^\prime$, $i^\prime$, $z^\prime$) at Lijiang station of Yunnan Observatories, UBAI-AZT-22 (0.91 d in $R$ band) at Maidanak Observatory, AC-32 telescope at Abastumani observatory (0.94 d in $R$), KAO (0.96 d in $g^\prime, r^\prime, i^\prime$) at Kottamia Observatory, Lisnyky-Schmidt (0.98 d in $R$) at Kyiv Observatory, NAO-50/70cm Schmidt (1.03 d in $I$) at Rozhen National Astronomical Observatory, CAHA (1.049 d in $g^\prime, r^\prime, i^\prime$) at Calar Alto Astronomical Observatory, T193/MISTRAL (1.050 d in $r^\prime$) at Haute-Provence Observatory, the 2-m telescope (1.08 d in $R$) at Shamakhy Astrophysical Observatory of Azerbaijan, FRAM-CTA-N (1.12 d in $R$) at Roque de los Muchachos Observatory, NOWT (2.09 d in $BVR$) at Xinjiang Astronomical Observatory, NOT (2.185 d in  $g^\prime, r^\prime, i^\prime$) at Roque de los Muchachos Observatory, NAO-2m (4.05 d in $r^\prime$, $i^\prime$) at Rozhen NAO\footnote{Using the focal reducer FoReRo-2 \citep{2000KFNTS...3...13J}.}, C2PU (10.10 d in $r^\prime$) at Calern observatory, CFHT-Megacam (30.48 d in $g^\prime, r^\prime, i^\prime$) at Mauna Kea Observatory. 

Near-infrared (NIR) observations of GRB~230812B were carried out with the Italian 3.6-m TNG telescope, sited in Canary Island, using the NICS instrument in imaging mode. A series of images were obtained with the J and K filters on 2023 August 16 (\textit{i.e.} about 4.1 days after the burst) and with the J filter only on 2023 August 21 and 2023 October 11 (\textit{i.e.} about 9.1 days and 60.1 days after the burst).

In addition to the professional network, GRANDMA activated its \href{http://kilonovacatcher.in2p3.fr/}{Kilonova-Catcher} (KNC) citizen science program for further observations with amateurs' telescopes.

The GRANDMA observations and its partners are listed in Table \ref{tab:all_observations}), which includes the start time $\mathrm{T_{mid}}$ time (in ISO format with post-trigger delay) and the host-galaxy/extinction-corrected brightness (in AB magnitudes) of the observations, as well as the uncorrected magnitudes. The exposure times, names of telescopes, and filters used are mentioned for each observation. Our method for calculating the magnitudes is described in the section 2.2, including our methods of photometry transient detection, magnitude system conversion, host galaxy extinction correction, and galaxy subtraction. 

\begin{table*}

\caption{X-ray and radio data used in this work. "Delay" is the time interval between the start of the observation ($T_{\rm start}$) and the \textit{Fermi} GBM's GRB trigger time (2023-08-12T18:58:12). We display both the unabsorbed flux densities and the corresponding computed AB magnitudes.}
\label{tab:Xandradio}
\scalebox{0.8}{

\begin{tabular}{|cc|cc|c|ccc|c|}
\hline
\multicolumn{2}{|c|}{$T_{\rm start}$}  & \multicolumn{2}{c|}{Delay} & Band & \multicolumn{3}{c|}{Flux} & Instrument  \\
UT & MJD & (day) & (s)  & Central frequency  & AB Magnitude & Flux density (Jy) & Error (Jy)  & \\
\hline
\multicolumn{9}{|c|}{X-ray bands}\\
\hline
2023-08-13T02:15:22 & 60169.094 & 0.304 & 26230 & 10 keV & 25.50 $\pm$ 0.45 & 2.29$\times10^{-7}$ & 9.5$\times10^{-8}$ & \textit{Swift} XRT \\
2023-08-13T03:48:00 & 60169.158 & 0.368 & 31788 & 10 keV & 25.78 $\pm$ 0.35 & 1.77$\times10^{-7}$ & 5.7$\times10^{-8}$ & \textit{Swift} XRT \\
2023-08-13T05:20:10 & 60169.222 & 0.432 & 37317 & 10 keV & 26.45 $\pm$ 0.37 & 9.55$\times10^{-8}$ & 3.3$\times10^{-8}$ & \textit{Swift} XRT \\
2023-08-15T00:20:43 & 60171.014 & 2.224 & 192151 & 10 keV & 28.01 $\pm$ 0.25 & 2.27$\times10^{-8}$ & 5.3$\times10^{-9}$  & \textit{Swift} XRT \\
2023-08-15T06:25:05 & 60171.267 & 2.477 & 214012 & 10 keV & 28.28 $\pm$ 0.26 & 1.77$\times10^{-8}$ & 4.2$\times10^{-9}$  & \textit{Swift} XRT \\
2023-08-18T22:40:51 & 60174.945 & 6.155 & 531759 & 10 keV & 30.15 $\pm$ 0.20 & 3.15$\times10^{-9}$ & 5.7$\times10^{-10}$  & \textit{Swift} XRT \\
2023-08-24T07:02:27 & 60180.293 & 11.503 & 993854 & 10 keV & 30.25 $\pm$ 0.23 & 2.89$\times10^{-9}$ & 6.1$\times10^{-10}$ & \textit{Swift} XRT \\
2023-08-29T05:47:45 & 60185.241 & 16.451 & 1421373 & 10 keV & 31.41 $\pm$ 0.59 & 9.94$\times10^{-10}$ & 5.4$\times10^{-10}$ & \textit{Swift} XRT \\
\hline
\multicolumn{9}{|c|}{Radio bands}\\
\hline
2023-08-14T18:13:52 & 60170.760 & 1.969 & 170139 & 15.5 GHz & 17.78 $\pm$ 0.16 & 2.8$\times10^{-4}$ & 4$\times10^{-5}$ & AMI-LA \\ 
2023-08-15T01:52:24 & 60171.078
 & 2.288 & 197652 & 6 GHz & 18.00 $\pm$ 0.05 & 2.3$\times10^{-4}$ &  1$\times10^{-5}$ & VLA\\ 
2023-08-15T01:52:24 & 60171.078
 & 2.288 & 197652 & 10 GHz & 18.17 $\pm$ 0.04 & 1.96$\times10^{-4}$ & 7$\times10^{-6}$ & VLA\\ 
2023-08-16T13:49:00 & 60172.576 & 3.785 & 327048 & 75 GHz & 18.55 $\pm$ 0.33 & 1.38$\times10^{-4}$ & 4.2$\times10^{-5}$ & NOEMA\\ 
2023-08-16T13:49:00 & 60172.576 & 3.785 & 327048 & 90 GHz & 18.87 $\pm$ 0.40 & 1.03$\times10^{-4}$ & 3.8$\times10^{-5}$ & NOEMA\\ 
2023-09-02T18:24:52 & 60189.767 & 20.977 & 1812399 & 3 GHz & 20.19 $\pm$ 0.40 & 3.06$\times10^{-5}$ & 1.12$\times10^{-5}$ & VLA\\ 
2023-09-02T18:24:52 & 60189.767 & 20.977 & 1812399 & 6 GHz & 19.67 $\pm$ 0.17 & 4.92$\times10^{-5}$ & 7.9$\times10^{-6}$ & VLA\\ 
2023-09-02T18:24:52 & 60189.767 & 20.977 & 1812399 & 6 GHz & 20.27 $\pm$ 0.39 & 2.82$\times10^{-5}$ & 1.01$\times10^{-5}$ & VLA\\ 
2023-09-17T11:30:00 & 60204.479 & 35.689 & 3083508 & 1.26 GHz &  $>$ 19.82 & $<$ 4.3$\times10^{-5}$ & - & uGMRT \\ 
\hline
\end{tabular}
}

\end{table*}

\begin{table*}
\caption{UVOIR observations of GRB 230812B. In column (2),  $T_{\rm (s)}$ is the time delay between the start of the observation and the \textit{Fermi} GBM's GRB trigger time (2023-08-12T18:58:12), all in days. Column (5) gives apparent magnitudes or 5-$\sigma$ upper-limits in the AB system, without any correction. Column (6) gives magnitudes in the AB systems for the afterglow and the associated SN, i.e. corrected for the host galaxy and the dust from the MW (AG + SN). When only upper limits were obtained, we corrected only for the MW dust. In Column (7), a cross means we did use this data point for the Bayesian analysis; in some cases the data were not used due to redundancy, i.e. a better measurement was made by another telescope at about the same time.} 

\label{tab:all_observations}
\scalebox{0.7}{
\begin{tabular}{|c|cc|c|c|c|c|c|c|}
\hline
\multicolumn{1}{|c|}{$T_{\rm start}$}  & \multicolumn{2}{c|}{$T_{\rm start}$ (days)} & Filter & Exposure & Magnitude & Corrected Magnitude & Telescope & Analysis  \\
\multicolumn{1}{|c|}{UT} & MJD & T-T$_{GRB}$ &  &  &  &  &  &  \\
\multicolumn{1}{|c|}{(1)} & \multicolumn{2}{c|}{(2)} & \multicolumn{1}{c|}{(3)} & \multicolumn{1}{c|}{(4)} & \multicolumn{1}{c|}{ Apparent (5)} & \multicolumn{1}{c|}{ AG + SN (6)} & & \multicolumn{1}{c|}{(7)}  \\
\hline


\hline
\multicolumn{9}{|c|}{$uv$ band} \\
\hline

2023-08-13T02:01:49  & 60169.085 & 0.294 & $white$ & - & 19.05 $\pm$ 0.03 &  18.97 $\pm$ 0.03 &  UVOT  & x \\
2023-08-13T03:33:56  & 60169.149 & 0.358 & $white$ & - & 19.36 $\pm$ 0.03 &  19.28 $\pm$ 0.04 &  UVOT  & x \\
2023-08-13T05:12:56  & 60169.217 & 0.427 & $white$ & - & 19.58 $\pm$ 0.04  &  19.51 $\pm$ 0.05  &  UVOT  & x\\
2023-08-15T00:12:30 & 60171.009 &  2.218 & $white$ & - & 21.62 $\pm$ 0.11 &  21.72 $\pm$ 0.21 &  UVOT & x\\
2023-08-17T12:14:49 & 60173.510 & 4.720 & $white$ & - & 22.77 $\pm$ 0.23  & 23.40 $\pm$ 0.88  &  UVOT & x\\
2023-08-19T04:17:04 & 60175.179 & 6.388 & $white$ & - & $>$ 22.82  &  $>$ 22.72  &  UVOT & x\\

\hline
\multicolumn{9}{|c|}{$g$ band} \\
\hline

2023-08-13T04:50:07 & 60169.201 & 0.411 & $g^\prime$ & 1$\times$300 s & 19.22 $\pm$ 0.10 & 19.14 $\pm$ 0.10 &  ZTF & x\\

2023-08-13T18:00:11 & 60169.750 & 0.960 & $g^\prime$ & 10$\times$180 s & 20.49 $\pm$ 0.08 & 20.47 $\pm$ 0.08 & KAO & x\\

2023-08-13T20:09:14 & 60169.840 & 1.049 & $g^\prime$ & 4$\times$240 s & 20.50 $\pm$ 0.03 & 20.48 $\pm$ 0.03 & CAHA &  \\

2023-08-14T12:47:47 & 60170.533 & 1.743 & $g^\prime$ & 1$\times$600 s & 21.00 $\pm$ 0.14 & 21.01 $\pm$ 0.15  & GMG & x\\

2023-08-14T23:41:38 & 60170.987 & 2.197 & $g^\prime$ & 1$\times$600 s & 21.6 $\pm$ 0.03 & 21.68 $\pm$ 0.04 & NOT & x \\

2023-08-15T20:48:26 & 60171.867 & 3.077 & $g^\prime$ & 4$\times$300 s & 22.18 $\pm$ 0.09 & 22.39 $\pm$ 0.10 & CAHA & x \\

2023-08-18 20:57:22 & 60174.873 & 6.083 & $g^\prime$ & 4$\times$300 s & 22.24 $\pm$ 0.3 & 22.46 $\pm$ 0.40 & CAHA & x \\

2023-08-19T19:35:17 & 60175.816 & 7.026 & $g^\prime$ & 14$\times$180 s & $>$ 21.2 & $>$ 21.1 & KAO & \\ 

2023-08-21T22:00:18 & 60177.917 & 9.126 &  $g^\prime$ & 3$\times$500 s &  22.83  $\pm$ 0.05 & 23.34 $\pm$ 0.12 & NOT & x \\

2023-09-04T15:03:01 & 60191.627 & 22.840 & $g^\prime$ & 2$\times$600 s & $>$ 21.8 & $>$ 21.7 &  GMG & \\

2023-09-12T6:38:16 & 60199.277 & 30.486 & $g^\prime$ & 10$\times$30 s & 23.32 $\pm$ 0.05 & 24.40 $\pm$ 0.27 &  CFHT-MegaCAM & x\\
\hline
\multicolumn{9}{|c|}{$r$ band} \\
\hline

2023-08-13T03:34:57 & 60169.149 & 0.359 & $r^\prime$ & 1$\times$300 s & 18.85 $\pm$ 0.05 & 18.80 $\pm$ 0.05 &  ZTF & x\\

2023-08-13T18:41:01 & 60169.778 & 0.989 & $r^\prime$ & 10$\times$180 s & 20.19 $\pm$ 0.11 & 20.24 $\pm$ 0.12 &  KAO & x\\

\hline
\end{tabular}}
\end{table*}

\begin{table*}
\addtocounter{table}{-1}
\caption{Continued.}
\scalebox{0.7}{
\begin{tabular}{|c|cc|c|c|c|c|c|c|}
\hline
\multicolumn{1}{|c|}{T$_{start}$}  & \multicolumn{2}{c|}{T$_{start}$ (days)} & Filter & Exposure & Magnitude & Corrected Magnitude & Telescope & Analysis  \\
\multicolumn{1}{|c|}{UT} & MJD & T-T$_{GRB}$ &  &  &  & & &    \\
\multicolumn{1}{|c|}{(1)} & \multicolumn{2}{c|}{(2)} & \multicolumn{1}{c|}{(3)} & \multicolumn{1}{c|}{(4)} & \multicolumn{1}{c|}{ Apparent (5)} & \multicolumn{1}{c|}{ AG + SN (6)} & & \multicolumn{1}{c|}{(7)}  \\
\hline

2023-08-13T20:09:15 & 60169.840 & 1.050 & $r^\prime$ & 4x300 s & 20.39 $\pm$ 0.05 & 20.46 $\pm$ 0.06 &  T193/MISTRAL & x\\

2023-08-13T20:23:43& 60169.850 & 1.059 & $r^\prime$ & 4$\times$240 s & 20.38 $\pm$ 0.02 & 20.45 $\pm$ 0.03 &  CAHA & \\


2023-08-14T12:58:38 & 60170.541 & 1.750 & $r^\prime$ & 1$\times$600 s & 20.98 $\pm$ 0.06 & 21.15 $\pm$ 0.08 & GMG & x\\

2023-08-14T23:24:41 & 60170.975 & 2.185 & $r^\prime$ & 1$\times$600 s & 21.28 $\pm$ 0.04 & 21.52 $\pm$ 0.06 & NOT & x\\


2023-08-15T16:36:09 & 60171.692 & 2.901 & $r^\prime$ & 6$\times$100 s & 21.60 $\pm$ 0.10 & 21.97 $\pm$ 0.15 &  NOWT & x\\

2023-08-15T21:12:19 & 60171.884 & 3.093 & $r^\prime$ & 3$\times$300 s & 21.78 $\pm$ 0.1 & 22.25 $\pm$ 0.17 &  CAHA & \\



2023-08-17T00:31:21 & 60173.0218 & 4.231 & $r^\prime$ & 18$\times$600 s & 21.78 $\pm$ 0.04 & 22.25 $\pm$ 0.09 &  NOT & x\\

2023-08-17T15:36:49 & 60173.651 & 4.860 & $r^\prime$ & 18$\times$200 s & 22.15 $\pm$ 0.15 & 22.93 $\pm$ 0.34 &  NOWT & x\\

2023-08-17T21:46:15 & 60173.907 & 5.117 & $r^\prime$ & 1$\times$1200 s & 21.85 $\pm$ 0.1 & 22.36 $\pm$ 0.18 &  T193/MISTRAL & x\\

2023-08-18T20:55:50 & 60174.872 & 6.081 & $r^\prime$ & 1$\times$600 s & 21.90 $\pm$ 0.04 & 22.45 $\pm$ 0.10 &  NOT & x\\

2023-08-18T21:22:09 & 60174.890 & 6.100 & $r^\prime$ & 3$\times$300 s & 21.71 $\pm$ 0.14 & 22.14 $\pm$ 0.22 &  CAHA & x\\



2023-08-19T17:11:42 & 60175.716 & 6.926 & $r^\prime$ & 15$\times$200 s & 21.59 $\pm$ 0.07 & 21.95 $\pm$ 0.11 &  NOWT & x\\


2023-08-21T16:41:57 & 60177.696 & 8.905 & $r^\prime$ & 16$\times$100 s & 21.55 $\pm$ 0.07 & 21.90 $\pm$ 0.11 &  NOWT & x\\

2023-08-21T18:37:20 & 60177.776 & 8.986 & $r^\prime$ & 18$\times$180 s & 21.40 $\pm$ 0.13 & 21.69 $\pm$ 0.18 &  KAO & x\\

2023-08-21T21:32:28 & 60177.898 & 9.107 &  $r^\prime$ & 1$\times$600 s & 21.76 $\pm$ 0.04 & 22.21 $\pm$ 0.09 &  NOT & x\\

2023-08-22T21:10:00 & 60178.882 & 10.092 & $r^\prime$ & 15$\times$300 s & 21.62 $\pm$ 0.06 & 22.00 $\pm$ 0.1 &  C2PU & x\\

2023-08-26T22:45:58& 60182.949 & 14.158 & $r^\prime$ & $1\times$600 s & 21.64 $\pm$ 0.09 & 22.03 $\pm$ 0.14 &  NOT & x\\ 




2023-09-04T15:23:48 & 60191.642 & 22.850 & $r^\prime$ & 2$\times$600 s & $>$ 21.7 & $>$ 21.65 &  GMG & \\

2023-09-05T14:33:44 & 60192.607 & 23.816 & $r^\prime$ & 16$\times$100 s & $>$ 21.9 & $>$ 21.85 &  NOWT & \\ 

2023-09-07T22:14:52 & 60194.927  & 26.136 & $r^\prime$ & $3\times$1800 s & 22.12 $\pm$ 0.03 & 22.86 $\pm$ 0.13 &  NOT & x\\ 


2023-09-09T19:30:00 & 60196.813 & 28.022 & $r^\prime$ & 5$\times$600 s & 22.35 $\pm$ 0.10 & 23.41 $\pm$ 0.33 &  NAO-2m & x\\

2023-09-11T19:14:55 & 60198.802 & 30.012 & $r^\prime$ & 11$\times$600 s & 22.18 $\pm$ 0.10 & 23.00 $\pm$ 0.26 & NAO-2m & \\

2023-09-12T06:31:48 & 60199.272 & 30.482 & $r^\prime$ & 3$\times$40 s & 22.33 $\pm$ 0.05 & 23.36 $\pm$ 0.22 &  CFHT-MegaCAM & x\\

2023-09-12T19:30:18 & 60199.813 & 31.023 & $r^\prime$ & 11$\times$600 s & 22.31 $\pm$ 0.10 & 23.31 $\pm$ 0.31 & NAO-2m & \\

2023-09-19T19:58:00 & 60206.832 & 38.042 & $r^\prime$ & 9$\times$600 s & 22.45 $\pm$ 0.1 & 23.72 $\pm$ 0.41 &  T193/MISTRAL & x\\

\hline

\multicolumn{9}{|c|}{$i$ band} \\
\hline

2023-08-13T19:28:23 & 60169.811 & 1.021 & $i^\prime$ & 2$\times$150 s & 20.16 $\pm$ 0.05 & 20.25 $\pm$ 0.06 &  KAO & x\\

2023-08-13T20:46:19 & 60169.865 & 1.705 & $i^\prime$ & 3$\times$240 s & 20.29 $\pm$ 0.03 & 20.40 $\pm$ 0.04 & CAHA & x\\

2023-08-14T13:09:36 & 60170.548 & 1.758 & $i^\prime$ & 1$\times$600 s & 21.08 $\pm$ 0.20 & 21.37 $\pm$ 0.27 &  GMG & x\\

2023-08-14T23:52:51 & 60170.995 & 2.205 & $i^\prime$ & 1$\times$300 s & 21.31 $\pm$ 0.06 & 21.69 $\pm$ 0.11 & NOT & x\\


2023-08-15T21:30:45 & 60171.896 & 3.106 & $i^\prime$ & 4$\times$300 s & 21.59 $\pm$ 0.11 & 22.14 $\pm$ 0.21 & CAHA & x\\


2023-08-16T20:12:30 & 60172.842 & 4.052 & $i^\prime$ & 9$\times$300 s & 21.64 $\pm$ 0.07 & 22.22 $\pm$ 0.16 &  NAO-2m & x\\

2023-08-17T00:42:36 & 60173.030 & 4.239 & $i^\prime$ & 1$\times$600s & 21.74 $\pm$ 0.06 & 22.41 $\pm$ 0.16 & NOT & x\\

2023-08-17T19:12:05 & 60173.800 & 5.001 & $i^\prime$ & 14$\times$300s & 21.69 $\pm$ 0.09 & 22.31 $\pm$ 0.19 &  NAO-2m & x\\

2023-08-18T21:07:04 & 60174.880 & 6.089 & $i^\prime$ & 1$\times$600 s & 21.68 $\pm$ 0.04 & 22.29 $\pm$ 0.12 & NOT & x\\
 
2023-08-18T21:39:40 & 60174.903 & 6.112 & $i^\prime$ & 4$\times$300 s & 21.64 $\pm$ 0.15 & 22.22 $\pm$ 0.19 & CAHA & \\

2023-08-19T19:50:40 & 60175.827 & 7.037 & $i^\prime$ & 20$\times$180 s & 21.95 $\pm$ 0.20 & 22.85 $\pm$ 0.51 & KAO & x\\

2023-08-20T18:01:44 & 60176.751 & 7.961 & $i^\prime$ & 29$\times$180 s & 21.51 $\pm$ 0.06 & 22.00 $\pm$ 0.12 &  KAO & x\\

2023-08-21T19:41:01 & 60177.820 & 9.030 & $i^\prime$ & 20$\times$180 s & 21.36 $\pm$ 0.06 & 21.77 $\pm$ 0.11 &  KAO & x\\

2023-08-21T22:27:57 & 60177.936 & 9.146 & $i^\prime$ & 1$\times$600 s & 21.58 $\pm$ 0.04 & 22.12 $\pm$ 0.11 & NOT & x\\

2023-08-22T19:30:17 & 60178.813 & 10.022 & $i^\prime$ & 20$\times$180 s & 21.49 $\pm$ 0.09 & 21.97 $\pm$ 0.16 &  KAO & x\\

2023-08-23T19:56:13 & 60179.831 & 11.040 & $i^\prime$ & 19$\times$180 s & 21.42 $\pm$ 0.07 & 21.86 $\pm$ 0.13 &  KAO & x\\


2023-08-26T22:57:10 & 60182.956 & 14.165 & $i^\prime$ & 1$\times$600 s & 21.42 $\pm$ 0.08 & 21.86 $\pm$ 0.14 & NOT & x\\

2023-08-28T19:20:25 & 60184.806 & 16.015 & $i^\prime$ & 26$\times$180 s & 21.40 $\pm$ 0.11 & 21.83 $\pm$ 0.18 &  KAO & x\\

2023-09-04T15:50:00 & 60191.660 & 22.870 & $i^\prime$ & 2$\times$600 s & 21.51 $\pm$ 0.24 & 22.00 $\pm$ 0.40 &  GMG & x\\


2023-09-12T06:24:15 & 60199.267 & 30.476 & $i^\prime$ & 3$\times$60 s & 21.79 $\pm$ 0.04 & 22.51 $\pm$ 0.14 & CFHT-MegaCAM & x\\

2023-10-07T19:44:24 & 60224.822 & 56.032 & $i^\prime$ & 1$\times$3000 s & 22.34 $\pm$ 0.04 & 24.07 $\pm$ 0.60 & NOT & \\
\hline
\multicolumn{9}{|c|}{$z$ band} \\
\hline

2023-08-14T13:20:38 & 60170.556 & 1.766 & $z^\prime$ & 1$\times$600 s & $>$ 20.5 & $>$ 20.47  & GMG &  \\

2023-08-14T23:30:23 & 60170.979 & 2.189 & $z^\prime$ & 1$\times$600 s & 20.98 $\pm$ 0.08 & 21.31 $\pm$ 0.12 & NOT & x \\

2023-08-21T21:43:40 & 60177.905 & 9.115 & $z^\prime$ & 1$\times$900 s & 21.68 $\pm$ 0.10 & 22.49 $\pm$ 0.26 & NOT & x \\
\hline
\multicolumn{9}{|c|}{$J$ band} \\
\hline

2023-08-16T21:33:58 & 60172.899 & 4.108 & $J$ & 20$\times$50 s & 21.00  $\pm$ 0.17 & 21.67 $\pm$ 0.43 & TNG & \\ 

2023-08-21T21:29:22 & 60177.895 & 9.105 & $J$ & 45$\times$60 s & 21.26  $\pm$ 0.27 & 22.23 $\pm$ 0.57 & TNG & \\ 

2023-10-11T20:15:22 & 60228.844 & 60.054 & $J$ & 45$\times$80 s & 21.82  $\pm$ 0.32 & - & TNG & \\ 
\hline

\multicolumn{9}{|c|}{$K$ band} \\
\hline

2023-08-16T20:49:28 & 60172.868 & 4.077 & $K$ & 30$\times$50 s & 21.41  $\pm$ 0.33  & 21.40 $\pm$ 0.33 with host & TNG & \\ 

\hline


\hline
\end{tabular}}
\end{table*}

\section{Spectroscopic observation details} 

We used OSIRIS+ \citep{2000SPIE.4008..623C} mounted on the 10.4m Gran Telescopio Canarias (GTC) telescope at Roque de los Muchachos Observatory in La Palma, Canary Islands, Spain, to observe the afterglow and supernova that follow GRB 230812B. The observation consisted of spectroscopy with an exposure time of 3x900s and grism R1000B, with a wavelength coverage between 3600 and 7800 AA. The first spectrum started at 21:37 UT, 1.110 days after the burst, while the on August 24, 2023 at 21.79 hours UT, ~12.12 days after the GRB detection, close to the peak of the supernova emission \citep{2023GCN.34597....1A,2023GCN.34409....1D}. 

The last two epochs were initially programmed to be obtained with larger spacing between epochs, but due to weather and telescope scheduling they ended up being rather close in time. The first epoch was obtained with a single grism, R1000B, covering the range between 3700 and 7880 Å. The second epoch included two grisms, R1000B and R1000R, this second one adding coverage between 5100 and 10100 Å to cover the full optical spectrum. 

\section{Skyportal}

To store, display, and annotate GRANDMA data products in a follow-up campaign, we use \texttt{SkyPortal} \citep{WaCr2019, Coughlin_2023}, a powerful database, API, and web application for time-domain astronomy. We use it for its capabilities of ingesting multi-messenger triggers from GCNs in real-time, from where network-cognizant observation plans are automatically generated using \texttt{gwemopt} \citep{Coughlin2018a}. It also enables automated ingestion of transients from the Transient Name Server (TNS) identified by surveys such as the Zwicky Transient Facility \citep{Bellm2019, Graham2019PASP}, the Panoramic Survey Telescope and Rapid Response System (Pan-STARRS) \citep{MoKa2012}, or the Asteroid Terrestrial-impact Last Alert System (ATLAS) \citep{ToDe2018}; it is from TNS that we retrieved the discovery photometry. We store photometry information, including flux and limiting magnitude measurements from follow-up observations by GRANDMA's telescopes within \texttt{SkyPortal}, supplementing the data imported from TNS, to create light curves. From the dedicated source page, easy access is provided to many other database services, such as \texttt{Vizier} \citep{Ochsenbein_2000}. Finally, \texttt{SkyPortal} is used for simplifying interactions with Bayesian inference frameworks such as the Nuclear physics and Multi-Messenger Astronomy framework \texttt{NMMA}~\citep{Dietrich:2020efo,Pang:2022rzc}, which we discuss more in the main text. Operations on \texttt{SkyPortal} are conducted and monitored by "shifters", members of the collaboration organized in teams every week, and divided into four daily slots of six hours each to accommodate timezone constraints while maintaining 24/7 coverage. Shifters look out for new candidates from surveys (particularly LIGO-Virgo-KAGRA, LVK) and new GCN events on the platform and report on associated Slack channels which candidates are to be followed up or not based on pre-defined criteria. The shifts are also organized using \texttt{SkyPortal}'s dedicated page in the form of a calendar. Shifters or members of telescope teams are expected to upload executed observations data either manually or programmatically using the API. GCN circular-like documents can even be automatically generated, ensuring consistent formatting of the results reported to the General Coordinates Network while reducing the possibility for human errors to be made.

\section{Comparing different astrophysical scenarios \texttt{NMMA}}
\label{comparedifferentscenarios}


As discussed in Section~\ref{NMMA}, using \texttt{NMMA}, we can quantitatively compare different astrophysical scenarios in a Bayesian framework. We have performed studies using various models and jet geometries. \texttt{NMMA} is able to perform joint Bayesian inference of multi-messenger events containing gravitational waves, GRB afterglows, SNe, or kilonovae. In addition to scenarios mentioned above, we also consider two kilonova models, \texttt{Bu2023Ye}~\citep{Anand:2023jbz} and \texttt{Ka2017}~\citep{2017}, to accompany the Top-hat model as possible explanations of the dataset. The log Bayes factor $\ln \mathcal{B}$ of various models relative to the Power-law+SN, which is the best-performing model to be introduced shortly, can be found in Table~\ref{tab:log_bayes_factorall}. The posterior of $\sigma_{\rm sys}$ is also shown in Table~\ref{tab:log_bayes_factorall}. The Power-law+SN has the lowest value of $\sigma_{\rm sys}$, thus signifying a better fit compared to other models.  Comparing the differences between the log Bayes factor of scenarios with and without a SN component, we conclude that the presence of a SN component is statistically supported from a Bayesian perspective.

\begin{table}
\begin{tabular}{| l | c  c |}
\hline
Scenario & log Bayes factor $\ln\mathcal{B}$ & $\sigma_{\rm sys} \ \rm [mag]$ \\
\hline
Power-law+SN & ref & $0.140^{+0.061}_{-0.050}$ \\
\hline
Top-hat & $-56.612 \pm 0.298$ & $0.693^{+0.133}_{-0.117}$ \\
Top-hat+SN & $-18.360 \pm 0.331$ & $0.282^{+0.081}_{-0.077}$ \\
Top-hat+Bu2023Ye & $-56.682 \pm 0.297$ & $0.696^{+0.126}_{-0.103}$ \\
Top-hat+Ka2017 & $-56.633 \pm 0.297$ & $0.696^{+0.125}_{-0.108}$ \\
Gauss & $-56.235 \pm 0.298$ & $0.690^{+0.122}_{-0.116}$ \\
Gauss+SN & $-17.356 \pm 0.331$ & $0.282^{+0.080}_{-0.073}$ \\
Power-law & $-56.292 \pm 0.296$ & $0.693^{+0.127}_{-0.111}$ \\
\hline
\end{tabular}
    \caption{The log Bayes factor $\ln\mathcal{B}$ and $\sigma_{\rm sys}$ value (median with $95\%$ credible interval) inferred for different models.}
    \label{tab:log_bayes_factorall}
\end{table}

\begin{figure*}
    \centering
    \includegraphics[width=0.45\textwidth]{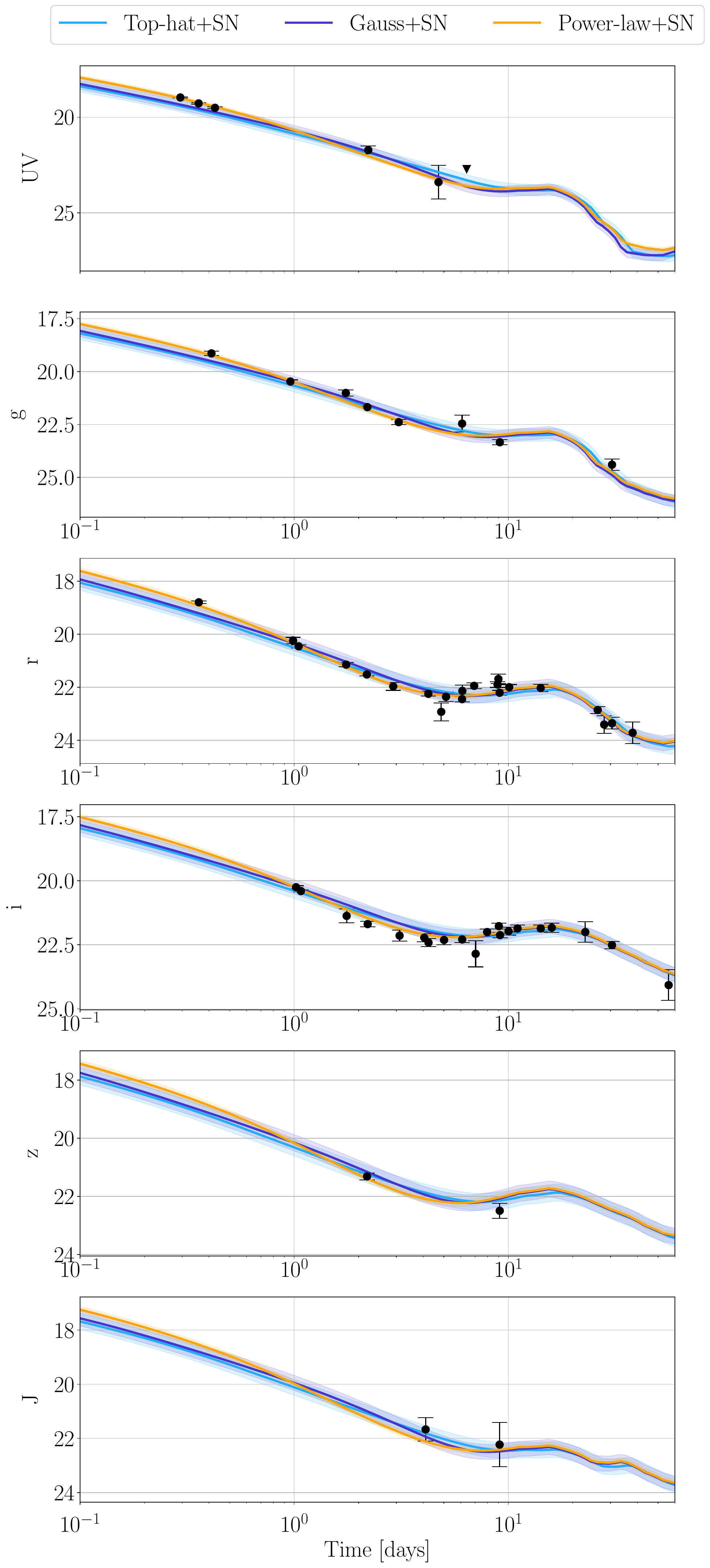}
    \includegraphics[width=0.45\textwidth]{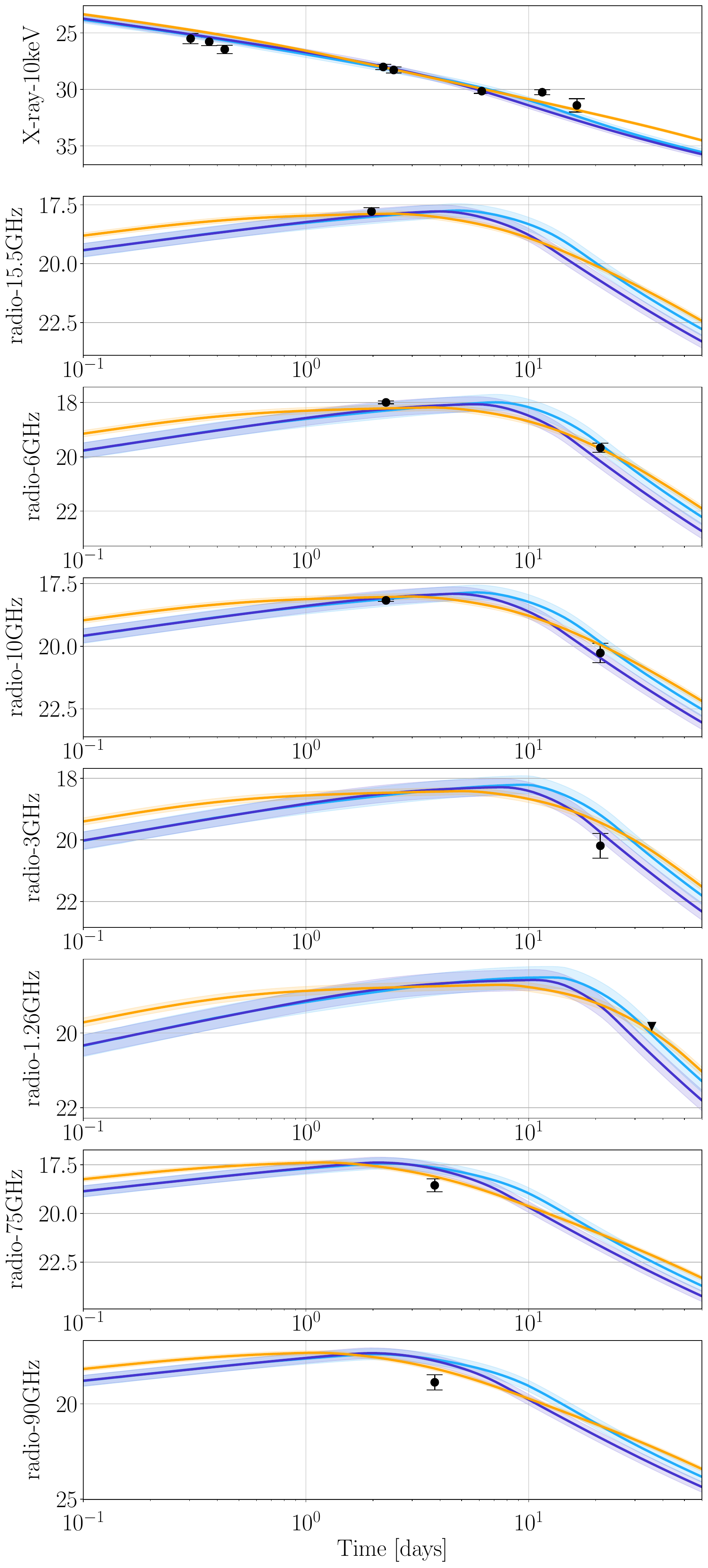}    
    \caption{Best-fit light curves for of the GRB + SN models with top-hat, Gaussian and power-law jet structures. Datapoints are reported in the observer frame.}
    \label{fig:bestfit_lcs_nmma}
\end{figure*}

The posterior values of all the considered models is summarized in Table~\ref{tab:MCMC_paramsNMMA_full}. The best-fit lightcurves for the three GRB + SN scenarios are shown in Figure~\ref{fig:bestfit_lcs_nmma}.

\begin{table}
    \renewcommand{\arraystretch}{1.1}
    \centering
    \caption{ \texttt{NMMA} - Parameters and prior bounds employed in our Bayesian inferences. We report median posterior values at 95~\% credibility for various physical scenarios and jet structures for the GRB. ``Uniform'' refers to an uniform distribution, and ``LogUniform'' refers to an uniform distribution for the $\log$ of the parameter. $\mathcal{N}(\mu, \sigma^2)$ refers to a Gaussian distribution with mean $\mu$ and variance $\sigma^2$.}
    \label{tab:MCMC_paramsNMMA_full} 
    \begin{turn}{90}
    \resizebox{0.7\paperheight}{!}{
\begin{tabular}{lcccccccccc}
\toprule
& & & \multicolumn{4}{c}{Top-hat} & & & &  \\ \cline{4-7}
Parameter & Prior & Prior range &  & +SN & +Bu2023Ye & +Ka2017 & Gauss & Gauss+SN & Power-law & Power-law+SN \\
\midrule
(log-) Isotropic afterglow energy $E_0$ [erg] & $\rm Uniform$ & $[47, 57]$ & $52.39^{\myplus0.40}_{\myminus0.44}$ & $52.91^{\myplus2.20}_{\myminus0.78}$ & $52.37^{\myplus0.38\phantom{0}}_{\myminus0.41\phantom{0} }$ & $52.39^{\myplus0.43\phantom{0}}_{\myminus0.39\phantom{0} }$ & $52.60^{\myplus0.68\phantom{0}}_{\myminus0.53\phantom{0} }$ & $52.58^{\myplus2.00\phantom{0}}_{\myminus0.42\phantom{0} }$ & $52.48^{\myplus0.45\phantom{0}}_{\myminus0.47\phantom{0} }$ & $52.82^{\myplus0.35\phantom{0}}_{\myminus0.31\phantom{0} }$ \\
(log-) Ambient medium's density $n_{\rm ism} $[$\rm{cm}^{-3}$] & $\rm Uniform$ & $[-6, 3]$ & $\myminus4.76^{\myplus1.36}_{\myminus1.23}$ & $\myminus1.36^{\myplus1.34}_{\myminus2.94}$ & $\myminus4.84^{\myplus1.21\phantom{0}}_{\myminus1.13\phantom{0} }$ & $\myminus4.89^{\myplus1.34\phantom{0}}_{\myminus1.10\phantom{0} }$ & $\myminus4.63^{\myplus1.65\phantom{0}}_{\myminus1.30\phantom{0} }$ & $\myminus1.38^{\myplus1.30\phantom{0}}_{\myminus2.27\phantom{0} }$ & $\myminus4.80^{\myplus1.37\phantom{0}}_{\myminus1.17\phantom{0} }$ & $\myminus2.38^{\myplus1.45\phantom{0}}_{\myminus1.60\phantom{0} }$ \\
(log-) Energy fraction in electrons $\epsilon_{\mathrm{e}}$ & $\rm Uniform$ & $[-5, 0]$ & $\myminus0.48^{\myplus0.39}_{\myminus0.38}$ & $\myminus0.13^{\myplus0.13}_{\myminus0.49}$ & $\myminus0.53^{\myplus0.39\phantom{0}}_{\myminus0.33\phantom{0} }$ & $\myminus0.56^{\myplus0.39\phantom{0}}_{\myminus0.33\phantom{0} }$ & $\myminus0.48^{\myplus0.42\phantom{0}}_{\myminus0.36\phantom{0} }$ & $\myminus0.12^{\myplus0.12\phantom{0}}_{\myminus0.43\phantom{0} }$ & $\myminus0.50^{\myplus0.40\phantom{0}}_{\myminus0.35\phantom{0} }$ & $\myminus0.10^{\myplus0.10\phantom{0}}_{\myminus0.29\phantom{0} }$ \\
(log-) Energy fraction in magnetic field $\epsilon_{\mathrm{B}}$ & $\rm Uniform$ & $[-10, 0]$ & $\myminus0.55^{\myplus0.47}_{\myminus0.86}$ & $\myminus4.02^{\myplus1.83}_{\myminus2.65}$ & $\myminus0.52^{\myplus0.45\phantom{0}}_{\myminus0.75\phantom{0} }$ & $\myminus0.52^{\myplus0.44\phantom{0}}_{\myminus0.80\phantom{0} }$ & $\myminus0.65^{\myplus0.56\phantom{0}}_{\myminus1.02\phantom{0} }$ & $\myminus3.56^{\myplus1.59\phantom{0}}_{\myminus2.52\phantom{0} }$ & $\myminus0.52^{\myplus0.44\phantom{0}}_{\myminus0.83\phantom{0} }$ & $\myminus2.29^{\myplus1.02\phantom{0}}_{\myminus0.94\phantom{0} }$ \\
Electron distribution power-law index $p$ & $\rm Uniform$ & $[2.01, 3]$ & $\phantom{0}2.13^{\myplus0.13}_{\myminus0.08}$ & $\phantom{0}2.25^{\myplus0.09}_{\myminus0.19}$ & $\phantom{0}2.14^{\myplus0.11\phantom{0}}_{\myminus0.09\phantom{0} }$ & $\phantom{0}2.15^{\myplus0.11\phantom{0}}_{\myminus0.09\phantom{0} }$ & $\phantom{0}2.13^{\myplus0.13\phantom{0}}_{\myminus0.08\phantom{0} }$ & $\phantom{0}2.15^{\myplus0.16\phantom{0}}_{\myminus0.09\phantom{0} }$ & $\phantom{0}2.13^{\myplus0.12\phantom{0}}_{\myminus0.08\phantom{0} }$ & $\phantom{0}2.04^{\myplus0.04\phantom{0}}_{\myminus0.02\phantom{0} }$ \\
Viewing angle $\theta_{\rm obs}$ [degrees] & $\mathcal{N}(0, \theta_{\rm core}^2)$ & -- & $\phantom{0}4.26^{\myplus9.18}_{\myminus4.26}$ & $\phantom{0}2.77^{\myplus2.24}_{\myminus2.42}$ & $\phantom{0}3.28^{\myplus6.28\phantom{0}}_{\myminus3.28\phantom{0} }$ & $\phantom{0}3.59^{\myplus6.36\phantom{0}}_{\myminus3.58\phantom{0} }$ & $\phantom{0}8.62^{\myplus11.51}_{\myminus8.61\phantom{0} }$ & $\phantom{0}2.20^{\myplus2.28\phantom{0}}_{\myminus2.11\phantom{0} }$ & $\phantom{0}7.33^{\myplus11.16}_{\myminus7.33\phantom{0} }$ & $\phantom{0}0.76^{\myplus1.29\phantom{0}}_{\myminus0.76\phantom{0} }$ \\
Jet core's opening angle $\theta_{\rm core}$ [degrees] & $\rm Uniform$ & $[0.6, 18]$ & $13.25^{\myplus4.74}_{\myminus6.35}$ & $\phantom{0}4.96^{\myplus3.30}_{\myminus4.20}$ & $13.20^{\myplus4.80\phantom{0}}_{\myminus5.75\phantom{0} }$ & $13.38^{\myplus4.61\phantom{0}}_{\myminus5.55\phantom{0} }$ & $11.02^{\myplus6.98\phantom{0}}_{\myminus7.25\phantom{0} }$ & $\phantom{0}3.96^{\myplus2.45\phantom{0}}_{\myminus3.09\phantom{0} }$ & $12.01^{\myplus5.98\phantom{0}}_{\myminus6.89\phantom{0} }$ & $\phantom{0}1.54^{\myplus1.02\phantom{0}}_{\myminus0.81\phantom{0} }$ \\
``Wing'' truncation angle $\theta_{\rm wing}$ [degrees] & $\rm Uniform$ & $[0.6, 45]$ & -- & -- & $22.28^{\myplus21.71}_{\myminus19.59}$ & $21.97^{\myplus20.71}_{\myminus20.47}$ & $27.83^{\myplus17.10}_{\myminus16.05}$ & $25.18^{\myplus19.77}_{\myminus17.50}$ & $29.21^{\myplus15.62}_{\myminus16.96}$ & $18.25^{\myplus19.07}_{\myminus11.48}$ \\
Power-law structure index $b$ & $\rm Uniform$ & $[0.1, 7]$ & -- & -- & $\phantom{0}3.48^{\myplus3.03\phantom{0}}_{\myminus3.36\phantom{0} }$ & $\phantom{0}3.46^{\myplus3.07\phantom{0}}_{\myminus3.36\phantom{0} }$ & -- & -- & $\phantom{0}3.46^{\myplus3.16\phantom{0}}_{\myminus3.27\phantom{0} }$ & $\phantom{0}1.64^{\myplus0.49\phantom{0}}_{\myminus0.47\phantom{0} }$ \\
Angle ratio $\theta_{\rm obs} / \theta_{\rm core}$ & -- & -- & $\phantom{0}0.33^{\myplus0.64}_{\myminus0.32}$ & $\phantom{0}0.56^{\myplus0.23}_{\myminus0.21}$ & $\phantom{0}0.27^{\myplus0.36\phantom{0}}_{\myminus0.27\phantom{0} }$ & $\phantom{0}0.29^{\myplus0.37\phantom{0}}_{\myminus0.29\phantom{0} }$ & $\phantom{0}1.03^{\myplus0.87\phantom{0}}_{\myminus1.03\phantom{0} }$ & $\phantom{0}0.64^{\myplus0.41\phantom{0}}_{\myminus0.57\phantom{0} }$ & $\phantom{0}0.68^{\myplus0.97\phantom{0}}_{\myminus0.68\phantom{0} }$ & $\phantom{0}0.49^{\myplus0.91\phantom{0}}_{\myminus0.49\phantom{0} }$ \\
Supernova boost $k_{\rm SN}$ & $\rm Uniform$ & $[0.01, 100]$ & -- & $\phantom{0}1.06^{\myplus0.18}_{\myminus0.18}$ & -- & -- & -- & $\phantom{0}1.08^{\myplus0.15\phantom{0}}_{\myminus0.15\phantom{0} }$ & -- & $\phantom{0}1.08^{\myplus0.09\phantom{0}}_{\myminus0.10\phantom{0} }$ \\
Supernova stretch $s_{\rm SN}$ & $\rm Uniform$ & $[0.1, 5.0]$ & -- & $\phantom{0}0.69^{\myplus0.08}_{\myminus0.07}$ & -- & -- & -- & $\phantom{0}0.68^{\myplus0.07\phantom{0}}_{\myminus0.07\phantom{0} }$ & -- & $\phantom{0}0.67^{\myplus0.05\phantom{0}}_{\myminus0.05\phantom{0} }$ \\
Systematic error $\sigma_{\rm sys}$ & $\rm LogUniform$ & $[0.01, 2.0]$ & $\phantom{0}0.69^{\myplus0.13}_{\myminus0.12}$ & $\phantom{0}0.28^{\myplus0.08}_{\myminus0.08}$ & $\phantom{0}0.70^{\myplus0.13\phantom{0}}_{\myminus0.10\phantom{0} }$ & $\phantom{0}0.70^{\myplus0.12\phantom{0}}_{\myminus0.11\phantom{0} }$ & $\phantom{0}0.69^{\myplus0.12\phantom{0}}_{\myminus0.12\phantom{0} }$ & $\phantom{0}0.28^{\myplus0.08\phantom{0}}_{\myminus0.07\phantom{0} }$ & $\phantom{0}0.69^{\myplus0.13\phantom{0}}_{\myminus0.11\phantom{0} }$ & $\phantom{0}0.14^{\myplus0.06\phantom{0}}_{\myminus0.05\phantom{0} }$ \\
\bottomrule
\end{tabular}
}
\end{turn}
\end{table}

\bsp	
\label{lastpage}
\end{document}